\documentclass[prd,superscriptaddress,nofootinbib,showpacs,preprint]{revtex4}
\usepackage{amsmath}
\usepackage{amsfonts}
\usepackage{amssymb}
\usepackage{graphicx}
	\usepackage{amsmath}
	\usepackage{IEEEtrantools}
	\usepackage[a4paper, total={6.5in, 8.5in}]{geometry}
\usepackage[colorlinks=true,linkcolor=blue,citecolor=blue]{hyperref}
	\usepackage{subfigure}
	
\begin{document}
\title{Exploring the Self Interacting Dark Matter Properties \\ From Low Redshift Observations}
\author{
Arvind Kumar Mishra 
${\href{https://orcid.org/0000-0001-8158-6602}{\includegraphics[height=0.15in,width=0.15in]{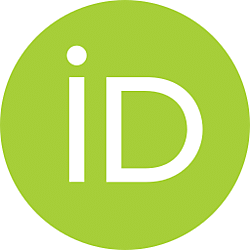}}}$~ 
}

\email{arvind@prl.res.in}

\affiliation{Theoretical Physics Division, Physical Research Laboratory,
	Navarangpura, Ahmedabad - 380009, India}
\affiliation{Indian Institute of Technology Gandhinagar, Palaj, Gandhinagar, 382355, India }

		\date{\today}
		
\def\be{\begin{equation}}
\def\ee{\end{equation}}

\begin{abstract}
The small scale observations indicate that the dark matter may be self-interacting. In this paper, we calculate the shear ($\eta$) and bulk viscosity ($\zeta$) of Self Interacting Dark Matter (SIDM) fluid, in kinetic theory formalism. Further, using the KSS bound on $\eta/\mathfrak{s}$, we derive a new upper limit on the ratio of dark matter self interaction cross section to its mass, $ \sigma/m $. Then, using the $ \sigma/m $ constraint, we show that KSS bound allows only sub-GeV mass of SIDM particle. Further, with the assumption of a  power-law form of $\eta$ and $\zeta$, we study its evolution in the light of low redshift observations. We find that at the large redshift, the SIDM viscosity is small but at the small redshift it becomes sufficiently large and contributes significantly in cosmic dissipation. As a consequence, viscous SIDM can explain the low redshift observations and also consistent with the standard cosmological prediction.
		\end{abstract}
		
%\pacs{12.38.Mh, 12.39.-x, 11.30.Rd, 11.30.Er}

		\maketitle
		\pagebreak
			
\section{Introduction}
The collisionless cold dark matter paradigm along with the cosmological constant ($\Lambda$CDM model of cosmology) explain the large scale structure of the Universe. But on the smaller scales, it faces major issues such as the core cusp problem, missing satellites problem, too big to fail problem, etc. For more detail on the small scale problems,
see reviews \cite{Weinberg:2013aya} \cite{Tulin:2017ara}.
It has been proposed that instead of the collisionless DM, if the DM particles interact with each other via elastic scattering over the scale where the problem is severe, then it can address the aforementioned problems \cite{Spergel:1999mh}\cite{Wandelt:2000ad}\cite{Randall:2007ph}\cite{Tulin:2013teo}\cite{Sarkar2018}.  The success of the SIDM lies in the fact that at the small scale due to large density, the
SIDM behaves like a collisional dark matter but on the large scale due to small density, it behaves like the collisionless DM. Thus the SIDM can  explain both the small and large scale observations very well. 

It is pointed out that the collisional nature of DM can lead to viscosity in DM fluid. In Ref.  \cite{Atreya:2017pny},  using the kinetic theory, we have calculated the viscous coefficients of the SIDM fluid. There we report that the 
theoretical constraint on the raio of the SIDM cross section to its mass ($\sigma/m$) for the present observed cosmic acceleration, is consistent with the constraint on $\sigma/m$ obtained from the astrophysical observations. Thus the viscous SIDM model can unify both the dark sectors, i.e. dark matter and the dark energy. 
Further, assuming the cluster size DM halo virialization and a power-law form of average velocity gradient, we study the cosmic evolution at the small redshifts in Ref. \cite{Atreya:2018iom}. There we find that decreasing velocity gradients can explain the low redshift cosmological observations. 

The inclusion of viscosity in the cosmic fluid has richer dynamical consequences in comparison with the ideal cosmic fluid. In literature, the effect of cosmic viscosity has been used in the different epochs of  cosmic evolution. It has been argued that the cosmic viscosity can explain the early time accelerated expansion (cosmic inflation)  \cite{Padmanabhan:1987dg, Gron:1990ew, Cheng:1991uu, Zimdahl:1996ka} and the late-time cosmic acceleration  \cite{Fabris:2005ts,Mathews:2008hk,Avelino:2008ph,Das:2008mj,  Piattella:2011bs, Velten:2011bg, Gagnon:2011id, Mohan:2017poq, Cruz:2018yrr,Li:2009mf,Barbosa:2015ndx,Floerchinger:2014jsa,Rezaei:2019ruy}. In Ref. \cite{Anand:2017wsj}, it has been argued that the viscous dark matter can reduce the tension between the Planck and local measurements of the Hubble expansion rate. Further, we show that the dark matter viscous dissipation can increase its temperature \cite{Bhatt:2019qbq}\cite{Mishra:2019uxl} and lead to visible photon production \cite{Mishra:2019uxl}.  These produced photons can increase the number density of photons in the Rayleigh-Jeans limit of the Cosmic Microwave Background (CMB) radiation and can explain the 21-cm anomaly reported by EDGES collaboration \cite{Mishra:2019uxl}. 
However, it has been found that large DM viscosity can increase the  DM temperature \cite{Bhatt:2019qbq}, decay of gravitational potential fluctuations \cite{Li:2009mf}, reduces the growth of the density perturbation \cite{Velten:2013pra}\cite{Velten:2014xca} and damping of the gravitational waves \cite{Goswami:2016tsu, Lu:2018smr,Brevik:2019yma}. These are severely constrained from the different astrophysical and cosmological observations. 
For recent work on cosmic viscosity, see Refs. \cite{Cai:2017buj,Anand:2017ktp,Medina:2019skp,Yang:2019qza,Bhatt:2019yld} and review \cite{Brevik:2017msy}.

In this work, we estimate the  bulk and shear viscosity of SIDM in kinetic theory framework and see its dependency  on the sound speed. We find  that for sufficiently large sound speed $ C_{n}>0.0027 $, $ \zeta $ becomes large in comparison with the $ C_{n}=0 $ case. Then, using the KSS bound, $ \eta/\mathfrak{s}\geq \frac{1}{4\pi}$ \cite{Kovtun:2004de}, we derive a new constraint on $ \sigma/m $, given by $ \frac{\sigma }{m}\leq \frac{(2\pi)^{\frac{5}{2}}}{(3)^{\frac{1}{2}}}\left( \frac{1}{m^{3}}\right) $. Further, we also estimate an upper limit on SIDM mass and show that KSS bound allowed only sub-GeV mass SIDM particle. 

We assume that on the small redshift  $ 0 \le z \le 2.5 $, the cluster scale may not be completely virialized but gravitationally bound and velocity gradient is constant on the cluster to a larger scale. In this case, the viscous coefficients of SIDM fluid may vary with the redshift. 
In order to study the viscous evolution, we consider the power-law form for bulk viscosity $\zeta(a)= \zeta_{0} \left(\frac{a}{a_{0}}\right) ^{\alpha}$ and shear viscosity  $\eta(a)= \zeta_{0} \left(\frac{a}{a_{0}}\right) ^{\alpha}$, where $ a $  and $ a_{0} $ represents the scale factor and its present value, respectively. We then use Einstein's equation and energy-momentum conservation equations and
calculate the background quantities such as the Hubble expansion rate, $H(z)$ and deceleration parameter, $q(z)$ for small redshift.  

Further, considering the length scale of spatial average $ \sim 20 $ Mpc \cite{Atreya:2018iom}, which is much larger than the cluster scale, we estimate the best fit value of the model parameter, $\alpha$ which explain the cosmic chronometer data and obtained the correct value of deceleration parmaeter in matter dominated era. The best fit value of model parameter suggests the decreasing DM viscosity on the earlier times (large redshift) and also explain the low redshift observations.
In our viscous SIDM fluid model, we also calculate the age of the Universe  and find that it is smaller than the age inferred from the CMB anisotropy data \cite{Tegmark:2006az} but larger than the globular cluster age  \cite{Carretta:1999ii}. 

The arrangement of our work is as follows: 
In Section \ref{sec:visccal}, we calculate the bulk and shear viscosity of SIDM from the kinetic theory and in relaxation time approximation. In Section \ref{etabysbound}, we derive a new upper limit on $ \sigma/m $ using the KSS bound on $ \eta/\mathfrak{s} $ and also estimate the constraint on SIDM mass. In Section \ref{sec:viscsidm}, using the Einstein field equations and energy-momentum conservation equations, we derive the dependency of the deceleration parameter on the SIDM viscosity. Further, in Section \ref{sec:vsidmcosm}, we derive the expression for mean free path and calculate the length scale over which the SIDM viscosity and spatial averages should be estimated. Then we set up the coupled differential equations for the Hubble rate and deceleration parameter. In Section \ref{sec:results} we estimate the best-fit values of the model parameter and  discuss our results. In the last Section \ref{sec:danc}, we conclude our work.

\section{Viscous Self-Interacting Dark matter (VSIDM)}
\label{sec:visccal}
In this section, we will calculate the SIDM viscosity in the framework of kinetic theory within the relaxation time approximation. Unlike the collisionless cold dark matter where only the gravitational interaction is important, in case the SIDM halo, the collisions between the DM particles also contribute to the DM halo formation. A large DM self-interaction causes the heat transfer between the inner and outer layer of the DM halo and causes the core profile towards the central region of the DM halo which matches the observations \cite{Ahn:2004xt}\cite{Rocha:2012jg}\cite{Peter:2012jh}.
As argued above that the observation on the small scales demands the non zero finite self-interaction between the DM particles, hence it is interesting to study its viscous effect. In this study, we are interested to calculate the viscous coefficients of SIDM fluid at a late time, when the DM halo has been gravitationally bound and more or less virialized.

In order to calculate the viscous coefficients of the SIDM, we apply the kinetic theory formalism in the relaxation time approximation. Using the kinetic theory and hydrodynamics, one can derive the expression for bulk $\zeta$ and shear viscosity $\eta$ as \cite{Gavin:1985ph, Kadam:2015xsa, Satapathy:2020sxs}
\begin{equation}
\label{eq:bulk}
\zeta = \frac{1}{T}\int\frac{d^3p}{(2\pi)^3}~\tau(E_{p})~\left[E_{p}C_{n}^2 - \frac{p^2}{3E_{p}}\right]^2 f_{p}^{0}~~,
\end{equation}
\begin{equation}
\label{eq:shear}
\mathrm{and\quad} \eta = \frac{1}{15T}\int \frac{d^{3}p}{(2\pi)^{3}}~\tau(E_{p})~\frac{p^{4}}{E_{p}^{2}}
~f_{p}^{0}~~,
\end{equation}  
where $ \tau(E_{p})$, $T$ and $ f_{p}^{0} $ represent the relaxation time, temperature and the equilibrium distribution function of the SIDM, respectively. Here $C_{n} = \frac{\partial P}{\partial \epsilon}\lvert_{n}$ is the speed of
sound at constant number density and $ E_{p} $ is the total energy of a SIDM particle. Further, one can also obtain the entropy density, $ \mathfrak{s} $ of the viscous SIDM medium as
\begin{equation}
\label{eq:entropy}
 \mathfrak{s} =\frac{1}{T} \int \frac{d^{3}p}{(2\pi)^{3}}~\left[E_{p}+ \frac{p^{2}}{E_{p}}\right] 
~f_{p}^{0}~~.
\end{equation} 
 In the relaxation time approximation, one assumes that collisions between the particles are sufficient enough to take system close to the local thermodynamic equilibrium in relaxation time. In this work, we approximate the relaxation time to the thermal average relaxation time, $ \tilde{\tau} $. For the scattering process $ a (p_{a})+b(p_{b})\leftrightarrow c (p_{c})+d(p_{d}) $, $ \tilde{\tau}_{a} $ is defined as
\begin{equation}
\label{eq:avt}
\tilde{\tau}_{a}^{-1} =\sum_{b}^{} n_{b} \langle\sigma_{ab} v_{ab}\rangle,
\end{equation}
where $n_{a}$ represents the number density of particle $a$, which is given by
\begin{equation}
n_{a} = \int_{0}^{\infty}\frac{d^{3}p_{a}}{(2\pi)^{3}}~f_{p_{a}}^{0}~~.
\end{equation}
Further, $ ~\langle\sigma_{ab} v_{ab}\rangle$ is the average
velocity weighted cross-section, defined as
\begin{equation}
\langle \sigma_{ab} v_{ab}\rangle =\frac{\int d^{3}p_{a}d^{3}p_{b}~\sigma~v_{ab}~\exp\left(\frac{-E_{a}}{T}\right) \exp\left(\frac{-E_{b}}{T}\right) }{\int d^{3}p_{a}d^{3}p_{b}~\exp\left(\frac{-E_{a}}{T}\right) \exp\left(\frac{-E_{b}}{T}\right)}
\end{equation}
We see that the value of $ \tilde{\tau}_{a}$ depends on the particle physics motivated model. Here we assume that the DM particles are colliding with themselves elastically and take $ n_{a}=n $
and $ \langle \sigma_{ab} v_{ab}\rangle=\langle \sigma v\rangle $.
We emphasize the assumption that the relaxation time approximation holds, when the dark matter particle scatters with each other at least one time in the DM halo formation time $ t_{\mathrm{halo}} $, i.e. $ t_{\mathrm{halo}}/\tilde{\tau}_{a}\approx 1 $. In Ref. \cite{Atreya:2017pny}, it has been shown that 
for the SIDM, the relaxation time approximation is valid at galactic and cluster scale and hence can be apply for its viscosity estimation.

In order to calculate the SIDM viscosity, we assume that the DM is non-relativistic, and follow the Maxwell-Boltzmann distribution. This implies that $ E_{p}\sim m+\frac{p^{2}}{2m} $ and $ f_{p}^{0}=\exp\left(-\frac{p^{\mu}u_{\mu}}{T} \right)$, where $ p^{\mu},u_{\mu} $ and $ m $ represents the four momentum, four velocity and mass of the SIDM particle.
In the rest frame of DM fluid and constant sound speed,  integration of Eq.(\ref{eq:bulk}) provides us an expression for the SIDM bulk viscosity as
\begin{equation}
\zeta  = \frac{m}{12\langle \sigma v\rangle}\left[ 12C^{4}_{n}\left(\frac{m}{T} \right)+ 5(4+9C^{4}_{n})\left(\frac{T}{m} \right) 
+ (-2+3C^{2}_{n})\left\lbrace 12C^{2}_{n}  + 70\left(\frac{T}{m} \right)^{2}\right\rbrace  + 315\left(\frac{T}{m} \right)^{3}\right] .
\label{eq:zetaspeed}
\end{equation}

Further, using the same assumptions as above, we can also estimate the shear viscosity  from Eq.(\ref{eq:shear}) as
\begin{equation}
\eta = \frac{m}{\langle \sigma v\rangle}\left(\frac{T}{m} \right)\bigg[1-7\left(\frac{T}{m} \right)+ \frac{63}{4}\left(\frac{T}{m} \right)^{2}\bigg]~.
\label{eq:etas}
\end{equation}
Further, from Eq.(\ref{eq:entropy}) the expression for entropy density is obtained as
\begin{equation}
\mathfrak{s}=\left( \frac{m^{2}}{2\pi}\right)^{\frac{3}{2}}\left(\frac{T}{m} \right)^{\frac{1}{2}} \left[2+5\left(\frac{T}{m} \right)-15\left(\frac{T}{m} \right)^{2} \right].
\label{eq:entropy1} 
\end{equation}
From above equations (\ref{eq:zetaspeed}), (\ref{eq:etas}) and (\ref{eq:entropy}), we see that the viscous coefficients ($ \eta, \zeta $) and entropy density depends on the ratio of SIDM temperature to its mass $( T/m )$ and its higher orders $ \mathcal{O} \left[  (T/m)^{2}, ...\right]  $.  We assume that the  VSIDM behaves like cold dark matter, hence $T/m \sim v^{2}_{\mathrm{vd}}\ll 1 $, where $ v_{\mathrm{vd}} $ is the DM velocity dispersion on the scale of our interest. The VSIDM model respect the DM coldness criteria since from cluster to supercluster scale $T/m$ varies from $ 10^{-5} $ to $ 10^{-4} $. 
Thus, for a good approximation it is sufficient to put the linear term in $T/m$ and neglect the higher order term in the expressions of Eq.(\ref{eq:zetaspeed}), Eq.(\ref{eq:etas}) and  Eq.(\ref{eq:entropy}). 
Therefore the simplified form of the bulk and shear viscosity is written as
\begin{equation}
\zeta =  \frac{m}{\langle \sigma v\rangle}\left(\frac{T}{m} \right)\left[ \frac{5}{3}\bigg( 1+\frac{9}{4}C^{4}_{n}\bigg) -2C^{2}_{n}\bigg(1-\frac{3}{2}C^{2}_{n}\bigg)\left( \frac{m}{T} \right) + C^{4}_{n}\left(\frac{m}{T} \right)^{2} \right],
\label{eq:zetas1}
\end{equation}
\begin{equation}
\mathrm{and} ~~~~ \eta = \frac{m}{\langle \sigma v\rangle}\left(\frac{T}{m} \right).
\label{eq:etas1}
\end{equation}
Here we find that the shear and bulk viscosities depend on the DM mass ($ m $), temperature $ T $, velocity average cross-section ($ \langle \sigma v\rangle $). We also point out that along with the above, $ \zeta $  also depends on sound speed, $ C_{n} $. We also stress that the above expression for shear and bulk viscosity is quite general and can be applied to the non-relativistic fluid which follow the Maxwellian distribution and validates the relaxation time approximation.

Further, considering the vanishing sound speed $C_{n} = 0 $ and using the argument of equi-partition of energy  $\frac{1}{2}m\langle v^{2} \rangle= \frac{3}{2}T$, we get the simplified form of the SIDM bulk and shear viscosity from Eq. (\ref{eq:zetas1}) and Eq. (\ref{eq:etas1}) as
\begin{equation}
\label{eq:viscosity}
\zeta =\frac{5.9}{9} \left( \frac{m}{\langle\sigma v\rangle}\right) \langle v\rangle^{2}~~~ \mathrm{and}~~~\eta = \frac{1.18}{3} \left( \frac{m}{\langle\sigma v\rangle}\right)\langle v\rangle^{2}~~.
\end{equation}
The above quoted form of the SIDM viscosity are the same as reported in earlier Ref. \cite{Atreya:2017pny}. Further, we  stress that for the rest of our calculations, we will use $ C_{n} =0 $, unless otherwise mentioned explicitly.
\section{Upper limit on $ \sigma/m $ and SIDM mass using $ \eta/\mathfrak{s} $ bound}
\label{etabysbound}
In this section, we will constraint the $ \sigma/m $ using the KSS bound on $\eta/\mathfrak{s}$ \cite{Kovtun:2004de}.  Then applying the $ \sigma/m $ constraint obtained from the numerical simulations, the upper limit on SIDM mass is derived. The ratio of shear viscosity to entropy density, $ \eta/\mathfrak{s}$ is obtained by using Eq. (\ref{eq:etas}) and (\ref{eq:entropy1}), which gives us
\begin{equation}
\frac{\eta}{\mathfrak{s}}=\frac{1}{2}\left(\frac{2\pi}{m^{2}} \right)^{\frac{3}{2}} \left( \frac{m}{\langle \sigma v \rangle}\right) \left( \frac{T}{m}\right)^{\frac{1}{2}}~.
\label{eq:etabys}
\end{equation}
From the above equation, we see that $ \eta/\mathfrak{s}$ depends on the velocity average cross section $ \langle \sigma v \rangle $, temperature, $ T $ and  mass of SIDM particle. 

In Ref. \cite{Kovtun:2004de}, Kovtun, Son and Starinets (KSS) have derived a universal lower bound on the ratio of shear viscosity to entropy density $ \eta/\mathfrak{s} $ (also known as KSS bound), given by 
\begin{equation}
\frac{\eta}{\mathfrak{s}}\geq \frac{1}{4\pi} ~~.
\label{eq:etabysbd}
\end{equation}
They argued that the lower bound on $ \eta/\mathfrak{s} $, i.e. $\frac{\eta}{\mathfrak{s}}= \frac{1}{4\pi} $ can be applied for various classes of quantum field theories.  

Here our interest is to derive a constraint on $\sigma/m$. This can be obtained by using Eq. (\ref{eq:etabys}) and Eq. (\ref{eq:etabysbd}), which provides us
\begin{equation}
\frac{\langle \sigma v \rangle}{m}\leq (2\pi)^{\frac{5}{2}}\left( \frac{1}{m}\right)^{3} \left( \frac{T}{m}\right)^{\frac{1}{2}}~.
\end{equation} 
Further, considering  $\langle \sigma v \rangle=\sigma \langle v \rangle  $ and $ v\sim \left(  \frac{3T}{m}\right)^{\frac{1}{2}}$, the above inequality simplifies as
\begin{equation}
\frac{\sigma }{m}\leq \frac{(2\pi)^{\frac{5}{2}}}{(3)^{\frac{1}{2}}}\left( \frac{1}{m}\right)^{3}~.
 \label{eq:sigmabymcontT}
\end{equation} 
This is a new upper limit on the ratio of SIDM self-interaction cross-section to its mass using the KSS bound. Further for a given $  \sigma/m  $ and using above Eq. (\ref{eq:sigmabymcontT}), we can also put a constraint on the DM mass,  given by 
\begin{equation}
m\leq\left[ \frac{\sqrt{3}}{(2\pi)^{\frac{5}{2}}} \left( \frac{m}{\sigma}\right)\right]^{\frac{1}{3}} ~.
\label{eq:dmmass}
\end{equation}
The above equation implies that the DM mass allowed from KSS bound depends only on $ \sigma/m $ . 
Therefore, we find that the constraint on DM properties such as its mass and self interacting cross section can be estimated once we calculate $ \sigma/m $ from other independent methods.

Further, we also point out that the explanation of small scale isssues demands another upper limit on $ \sigma/m $ of SIDM. Note that in this work, we are interested on cluster scale, thus we will use the cluster scale constaint on $ \sigma/m $. A strong upper limit on the cluster scale comes from the merging cluster IE 0657-56, which demands \cite{Randall:2007ph} 
\begin{equation}
\frac{\sigma }{m}< 1.25 \mathrm{cm}^2/\mathrm{gm}~.\label{eq:sigmabymcontO}
\end{equation} 
Further a velocity dependent cross section, which explain the cluster scale issues requires $ \sigma/m \approx0.1$cm$^{2}/$gm  \cite{Kaplinghat:2015aga}. 
So in order to get a constraint on the DM mass, the above two discussed values of $ \sigma/m $ will be used.
\begin{figure}[]
	\includegraphics[width=0.6\linewidth]{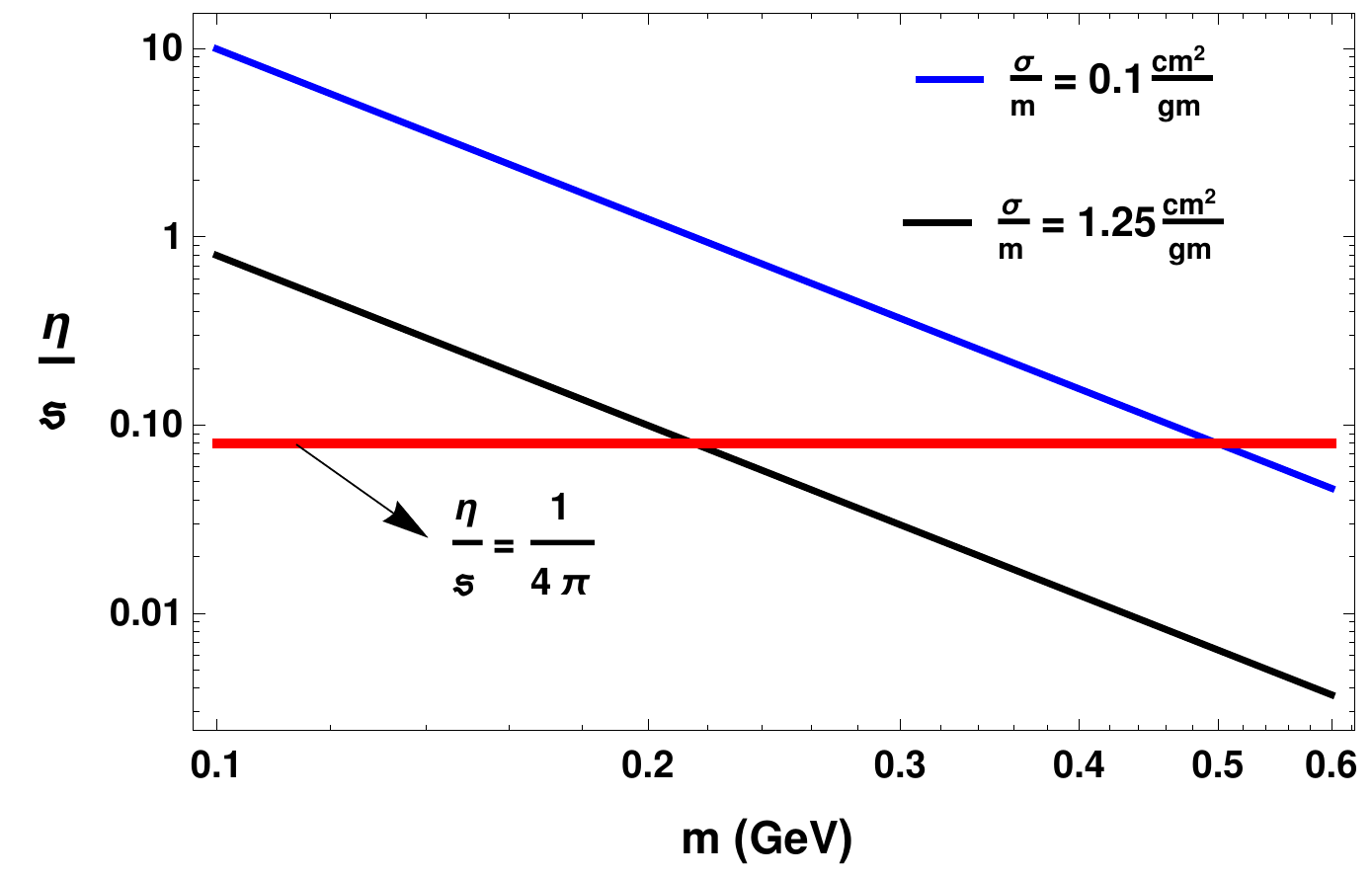}	
\caption{In Fig. \ref{fig:shearconst}, shear viscosity to entropy ratio, $ \eta/\mathfrak{s}$ is plotted as a function of DM mass. The red line corresponds for KSS lower bound, $\frac{\eta}{\mathfrak{s}} = \frac{1}{4\pi}$  \cite{Kovtun:2004de}.} 	
	\label{fig:shearconst}
\end{figure}	 

In Fig.\ref{fig:shearconst}, using Eq. (\ref{eq:etabys}), we plot $ \eta/\mathfrak{s} $ on the cluster scale as a function of DM mass, for different values of $ \sigma/m $ at present $ z=0 $. Here  solid blue and dashed blue lines corresponds for $\sigma/m =1.25$cm$^{2}/$gm and $\sigma/m =0.1 $cm$^{2}/$gm, respectively. The red solid line corresponds for the lower KSS bound, i.e. $ \eta/\mathfrak{s} = 1/4\pi$. 
Since $\eta/\mathfrak{s}\propto m^{-3}$, so it decreases for large DM mass, so it is possible that for some large DM mass $ \eta/\mathfrak{s}$ may be smaller than its lower limit inferred from KSS bound (red solid line). From Fig. \ref{fig:shearconst}, it is clear that larger will be the $\sigma/m $, smaller will be the DM mass allowed.
For $\sigma/m =1.25$cm$^{2}/$gm (upper limit from the cluster merger) , $m\leq 0.21 $ GeV and for  $\sigma/m =0.1$cm$^{2}/$gm (required to explain the cluster scale issues), we get $m\leq 0.5$ GeV. 

Thus we conclude that the KSS bound allowed only sub-GeV SIDM mass. As discussed above that the bound on  DM mass only depends on $ \sigma/m $ and will be improved for more precise estimation of $\sigma/m $. The direct detection experiments for the DM searches  have almost failed to detect the particle dark matter candidate of mass varies from GeV to TeV scale. Thus, our result can provide a new DM mass range, which will be important for future dark matter searches experiment. We refer Ref. \cite{Lin:2019uvt} for details of the experiment which will probe the sub-GeV DM mass range.
\section{Einstein equation in presence of the VSIDM}\label{sec:viscsidm}
In this Section, we investigate the effects of VSIDM fluid on the cosmic evolution history of the Universe through Einstein's equation and energy-momentum conservation equations following the formalism discussed in Refs. \cite{Floerchinger:2014jsa,Atreya:2017pny}.

In our VSIDM model, we assume that the Universe is mainly dominated by the dark matter with no dark energy component at present.
In the Landau frame with first-order gradient expansion, the energy-momentum tensor of the VSIDM can be written in terms of the ideal and the viscous contributions as
 \begin{equation}
T^{\mu\nu} =  T^{\mu\nu}_{\mathrm{ideal}} + T^{\mu\nu}_{\mathrm{visc}}
 \end{equation}
 where, the ideal and viscous terms are given by
  \begin{equation*}
 T^{\mu\nu}_{\mathrm{ideal}} = \epsilon u^{\mu} u^{\nu} + P  \Delta^{\mu \nu} 
  \end{equation*}
  and
   \begin{equation}
T^{\mu\nu}_{\mathrm{visc}}= \Pi_{B} \Delta^{\mu \nu} + \Pi^{\mu \nu}~,
   \end{equation}
where $\epsilon$, $ u^{\mu} $ and $P$ represents the energy density, four velocity and kinetic pressure of the dark matter fluid. The $ \Delta^{\mu \nu} = u^{\mu} u^{\nu} + g^{\mu \nu}$ is defined as the projection operator, which is orthogonal to the four velocity, i.e. The $ u_{\mu}\Delta^{\mu \nu}=0$. Here the $\Pi_{B}$ and $\Pi^{\mu \nu}$ represents bulk and shear stress which forms are given as
 \begin{equation}
 \Pi_{B} = - \zeta \nabla_{\mu} u^{\mu}
 \end{equation}
 \begin{equation}
 \Pi^{\mu \nu} =
  - \eta\left[ \Delta^{\mu\alpha}\Delta^{\nu\beta} + \Delta^{\mu\beta}\Delta^{\nu\alpha} - \frac{2}{3}\Delta^{\mu\nu}\Delta^{\alpha\beta}\right] \nabla_{\alpha}u_{\beta}~~.
 \end{equation}
 Here the shear stress is orthogonal to the four velocity, $u_{\mu} \Pi^{\mu \nu} = 0$ and traceless, $ \Pi_{\mu}^{\mu} = 0$.
 
To study the cosmic evolution in the VSIDM model, we apply the Einstein equations, $ G_{\mu\nu} =-8\pi G T_{\mu\nu}$ and energy momentum conservation equations, $ \nabla_{\mu} T^{\mu \nu} = 0. $ Further, we assume the scalar metric perturbation of the form
\begin{equation}
ds^{2} = a^{2}(\tau)\left[-\big( 1+2\psi(\tau,\vec{x})\big)  d\tau^{2} + \big( 1-2\phi(\tau,\vec{x})\big)  \right] 
\label{eq:metric}
\end{equation}
and neglect the vector and tensor perturbations in metric. In the above Eq. $ a(\tau) $ is scale factor and $ \psi, \phi$ are the potentials, respectively. At the late time, we assume $ \psi, \phi\ll1$  and fluid velocity is small, i.e. ${v}^{2}\ll 1$. Then, using the average energy density equation and average trace of Einstein's equation, one can get
the evolution equation of deceleration parameter $q$ as \cite{Floerchinger:2014jsa} 
\begin{equation}
\frac{dq}{dz} +  \left( \frac{q-1}{1+z}\right)  \bigg[2q-(1+ 3 \hat{w}_{\mathrm{eff}})\bigg] = \frac{4\pi G D}{3(1+z) H^{3}} \bigg(1 - 3 \hat{w}_{\mathrm{eff}}\bigg) \label{eq:qe}
\end{equation}
where,
	\begin{equation}
	\label{eq:diss}
D = (1+z)^{2}\left<\eta\left[\partial_{i}v_{j}
	\partial_{i}v_{j}+\partial_{i}v_{j}\partial_{j}v_{i} -
	\frac{2}{3}\partial_{i}v_{i}\partial_{j}v_{j}  \right]\right>_{s}
	+ (1+z)^{2}\left<\zeta[\vec{\nabla}\cdot\vec{v}]^2\right>_{s} +
(1+z)\left<{\vec{v}}\cdot\vec{\nabla}(P-6\zeta H)\right>_{s}.
	\end{equation}
Where $\langle A \rangle_{s}$ represents the spatial average of $A$. Here 
$v$ is the peculiar velocity and spatial derivative represents the derivative w.r.t. the comoving coordinate.
The effective Equation of State (EoS) is given by 
$ \hat{w}_{\mathrm{eff}}=\frac{\langle P\rangle_{\mathrm{eff}}}{\langle\epsilon\rangle_{s}}~$,
where $ \langle P\rangle_{\mathrm{eff}} $ is effective pressure, which is the sum of the kinetic and the bulk viscous pressure of SIDM, i.e. $\langle P\rangle_{\mathrm{eff}} = \langle P\rangle_{s} +\langle\Pi_{B}\rangle_{s} $~.
We also define the  EoS corresponding to the bulk viscous pressure of the VSIDM as $ \hat{w}_{B} =\frac{\langle\Pi_{B}\rangle_{s}}{\langle\epsilon\rangle_{s}}$~.
From Eq.(\ref{eq:qe}) 
it is clear that the dynamics of the Universe depends on the dark matter effective EoS, $ \hat{w}_{\mathrm{eff}} $ and $D$. 

Further, from Eq. (\ref{eq:qe}) it is manifest that the Universe will be in accelerated phase if \cite{Floerchinger:2014jsa} 
\begin{equation}
\frac{4\pi G D}{3 H^{3}} > \frac{1+3\hat{w}_{\mathrm{eff}} }{1-\hat{w}_{\mathrm{eff}}}~.
\label{eq:ac}
\end{equation}
provided that $ \hat{w}_{\mathrm{eff}}\neq 1 $. The above condition suggests that for an accelerated expansion, the dissipational effects of the DM should be sufficiently large. For example, for the present observed cosmic acceleration, $ \frac{4\pi G D}{3 H^{3}}|_{z=0} = 4.13 $ \cite{Atreya:2018iom}. Thus, in order to study the strength of viscous dissipation and its effect on the cosmic evolution, we need to calculate the $D$ term given in Eq.(\ref{eq:diss}), which will be done in the next Section.
\section{Viscous self interaction dark matter Cosmology}
\label{sec:vsidmcosm}
From the previous Section, we find that the dynamics of the Universe depends on $D$, which needs to be calculated. In this section, we will estimate the typical scale of viscosity and spatial average. Further, using the simplified assumptions, we will estimate $D$. Later we will set up the differential equations for the Hubble expansion rate and the deceleration parameter to study the effect of the SIDM viscosity on the cosmic evolution.
\subsection{Scale of the SIDM viscosity and spatial averages}
In this subsection, we will estimate the length scales on which the VSIDM viscosity and the spatial average should be calculated. The minimum scale on which the viscosity coefficients of VSIDM fluid can be calculated will depend on the length scale over which the fluid description of the SIDM particle is valid. The hydrodynamic description of the SIDM particles will be valid when the mean free path of the dark matter particle, $\lambda_{\mathrm{SIDM}}$ is less than the length scale under consideration, i.e. $\lambda_{\mathrm{SIDM}} < L$. In the dilute gas approximation, the mean free path of SIDM particle is given by $ \lambda_{\mathrm{SIDM}}= \frac{1}{\sqrt{2}}\left(\frac{1}{n\sigma}\right)$ \cite{book:1124099}. In a simplified manner, the expression for SIDM can be rewritten as
\begin{equation}
\lambda_{\mathrm{SIDM}}\sim 3\times  10^{9} \left(\frac{m/\sigma}{\mathrm{gm/cm}^{2}} \right) \left(\frac{ \mathrm{M}_{\odot}\mathrm{kpc}^{-3}}{\rho} \right) \mathrm{kpc}~,
\end{equation}
where $m/\sigma $ and $ \rho $ should be taken in units of $ \mathrm{gm/cm}^{2} $ and $ \mathrm{M}_{\odot}\mathrm{kpc}^{-3} $, respectively. The expression for $ \lambda_{\mathrm{SIDM}} $ calculated above is less than an order of the magnitude in comparison with the mean free path of the SIDM particles as given in Ref. \cite{Atreya:2017pny}. Using the isothermal profile for cluster scale and $ \rho\sim 2\times 10^{7}~\mathrm{M}_{\odot}\mathrm{kpc}^{-3}$,  $\sigma/m = 0.1$ cm$^2/$g as obtained from Ref. \cite{Kaplinghat:2015aga}, we get $ \lambda_{\mathrm{SIDM}}\sim 1$ Mpc, which is the order of the cluster scale ($\sim$ Mpc). Thus we assume that the length scale, where viscosity estimation has to be done, should be at least cluster or larger scales (i.e. supercluster scale), see also Refs. \cite{Atreya:2017pny}\cite{Atreya:2018iom}. 

Further, it is important to note that the length scale over which the spatial averaging should be done will be at least a cluster scale. As a result, all the velocity gradients should be estimated on the length scale larger to cluster size DM halo. Up to this discussion, a spatial averaging length scale of $L$ is a free parameter that will be calculated in the next Section.
\subsection{Calculation of  $D$ term}
From equation  (\ref{eq:diss}), we see that the term $D$ crucially depends on the spatial average of the velocity gradient and viscosities. Thus its calculation demands the explicit estimation of the spatial averages. In this work, without going into the detail calculation of the spatial averages, we will approximate the dissipation term $D$ using the following assumptions:

$(i)$ We assume that on the redshift of our interest $ 0\le z \le 2.5$, the length scale equal to or larger than the cluster scale, the spatial average peculiar velocity gradient $\langle\partial v \rangle_{s} $ is constant, i.e. $\langle\partial v \rangle_{s} \sim constant$. Then we can approximate 
\begin{equation}
\langle\partial v \rangle_{s} \sim v_{0}/L~~,
\label{eq:vbyl}
\end{equation}
where $v_{0}$ and $L$ represents the peculiar velocity and comoving length scale. Qualitatively our assumption implies that the peculiar velocity and  comoving length scale varies in such a way that so that the ratio of these two becomes constant for the low redshift. To check the validity of the above assumption at present, we calculate the $  v_{0}/L $ on typical cluster and supercluster scale and compare with each other. For typical cluster scale $L \sim $ Mpc, DM velocity $ v_{0} \sim 1000$ km$/$sec, so $ v_{0}/L \sim 10^{-18}$ sec$^{-1}$ and for typical supercluster scale  $L \sim 50$ Mpc and $v_{0}\sim6000$ km$/$sec so $ v_{0}/L\sim 10^{-18}$ sec$^{-1}$. Hence we find that at present, i.e. $ z=0 $, the velocity gradient, $  v_{0}/L $ is approximately same on both the cluster and supercluster scale and hence validates our assumption. 

$(ii)$
The variation of viscosity coefficients $\zeta,\eta$ are independent with space since they depend on the thermal distribution of the dark matter \cite{Atreya:2018iom}. We assume that the cluster may not be completely virialized  but gravitationally bound between the redshift of our interest $ 0 < z \le 2.5$ but at present, $ z=0 $, the clusters are more or less virialized. Thus, in this case, the dark matter viscosity may vary within the low redshift interval $ 0 < z \le 2.5$.

In order to study the variation in the viscous coefficient of $\zeta$ and $\eta$ from equation (\ref{eq:viscosity}) ($ C_{n} = 0 $), we need to understand the evolution of the $ m/\langle\sigma v\rangle $ and $ \langle v\rangle$ with redshift individually, that depends on the particle physics model of the SIDM.
In this study, attributing the redshift dependence information in the power law form, we consider the bulk and shear viscosities of the SIDM fluid as
\begin{equation}
\zeta(z) =\zeta_{0} \left(\frac{a}{a_{0}}\right) ^{\alpha}= \zeta_{0}\left(\frac{1}{1+z}\right) ^{\alpha} ,
\label{eq:zetaf}
\end{equation}
\begin{equation}
\mathrm{and} ~~~~\eta(z) =\eta_{0} \left(\frac{a}{a_{0}}\right) ^{\alpha}= \eta_{0}\left(\frac{1}{1+z}\right) ^{\alpha} ,
\label{eq:etaf}
\end{equation}
where $\alpha $ is the viscosity parameter and $ \frac{a}{a_{0}} =\frac{1}{1+z}$, where $ a_{0}=1 $ is the present value of the scale factor. Here $ \zeta_{0}=\zeta(z=0) $ and $ \eta_{0}=\eta(z=0) $ represents the present values of bulk and shear viscosity, respectively. Their values can be estimated from the equation (\ref{eq:viscosity}) at the cluster scale and can be given in term of the astrophysical units as
 \begin{equation}
 \zeta_{0}=6.6\times 10^{9}\left[  \frac{{(\rm{cm^{2}/gm) (km/ sec)}}}{\langle \sigma_{c}v_{c} \rangle/m}\right]\left(\frac{\langle v_{c} \rangle}{1000~\rm{km/ sec}} \right)^{2}~\mathrm{Ps-sec}~~,
 \label{eq:bulkpresent}
 \end{equation}  
  \begin{equation}
\mathrm{and} ~~~~ \eta_{0}=6\times 10^{9}\left[  \frac{{(\rm{cm^{2}/gm)(km/ sec)}}}{\langle \sigma_{c}v_{c} \rangle/m}\right]\left(\frac{\langle v_{c} \rangle}{1000~\rm{km/ sec}} \right)^{2}~\mathrm{Ps-sec}~~.
  \label{eq:shearpresent}
  \end{equation}  
Here $ \langle \sigma_{c}v_{c} \rangle=\langle \sigma v \rangle_{z=0} $ and  $ \langle v_{c}\rangle=\langle  v \rangle_{z=0}~$. 
We take cluster scale velocity,  $ \langle v_{c}\rangle \sim 1000$ km$/$sec and cluster scale 
$\frac{\langle \sigma_{0}v_{c} \rangle}{m} \sim 100$
(cm$^{2}$/gm)(km/sec) from Ref. \cite{Kaplinghat:2015aga}.  

Thus, using the above approximations, we may estimate $D$ as given in Eq. (\ref{eq:diss}) as 
\begin{equation}
D \sim 2\times 10^{-30} \left[  \frac{{(\rm{cm^{2}/gm) (km/ sec)}}}{\langle \sigma_{c}v_{c} \rangle/m}\right]\left(\frac{\langle v_{c} \rangle}{1000~\rm{km/ sec}} \right)^{2}\left(\frac{v_{0}}{1000~\rm{km/ sec}}\right)^{2}\left(\frac{\rm{Mpc}}{L} \right)^{2} \big(1+z\big)
^{2-\alpha}
\label{eq:Dassume}
\end{equation}
in the units of $ \rm{Ps}-\mathrm{sec^{-1}}$. 
From the above equation, it is clear that for very large averaging length scale ($ L \rightarrow \infty $) or vanishing viscosity coefficients ($ \eta=0, \zeta=0 $), $D$ term becomes zero and the VSIDM fluid behaves like a dissipationless fluid. We also see that for $\alpha=2$, the dissipation will constant with the redshift. Since, as discussed in the above assumption that the $ v_{0}/L $ is constant, $\eta_{0}$ and $\zeta_{0}$ is fixed at $z=0$, hence the evolution of $D$ will depend only on the  parameter, $\alpha$.
\subsection{ Background cosmology in VSIDM model }
\label{sec:backcosmo}
In this subsection, equipped with the simplified form of $D$, we will set up the equations for the Hubble expansion rate and deceleration parameter. Assuming the cold VSIDM fluid, i.e. $ \hat{w}_{\mathrm{eff}} \approx 0 $ and using Eq. (\ref{eq:Dassume}), Eq. (\ref{eq:qe}) simplifies as
\begin{equation}
\frac{dq}{dz} +  \frac{(q-1) \left(2q- 1\right)}{(1+z)} = \beta \left( \frac{1+z }{\bar{H}^{3}}\right), 
\label{eq:qev}
\end{equation}
where $ \beta  $ is dissipation parameter, which is given as
\begin{equation}
\beta =  \frac{4\pi G }{3H^{3}_{0}}\left(\frac{4}{3}{\eta_{0}}  +  2\zeta_{0} \right) \left(\frac{v_{0}}{L}\right)^2\big(1+z\big) ^{-\alpha}~. 
\label{eq:beta}
\end{equation}
where $\bar{H} = \frac{H}{H_{0}}$ is dimension-less Hubble expansion rate and $H_{0}=H(z=0)$ is the value of the present Hubble expansion rate. The dissipation term, $\beta$ depends on exponent power, $\alpha$ and averaging length scale, $L$.

In order to solve the Eq. (\ref{eq:qev}), we need to provide the expression for the Hubble expansion rate. For this purpose, we use the definition of deceleration parameter $q(z) = -1 + (1+z)~\frac{H'}{H} $  and obtain the differential equation for $ \bar{H} $ as
\begin{equation}
\frac{d\bar{H}}{dz} =  \left( \frac{q+1}{1+z}\right)\bar{H}~.
\label{eq:hevz}
\end{equation}
Thus, we have obtained the coupled differential equations in $ q(z)$ and $ \bar{H}(z)$ given by Eqs. (\ref{eq:qev}) and (\ref{eq:hevz}). These equations can be solved numerically by using the initial conditions   
at present, i.e.  $\bar{H}(z=0) = 1$ and $q(z=0) = -0.60$ \cite{Ade:2013zuv}. The solution of $ q(z) $ and $ \bar{H}(z)$ effectively depends on two free parameters $ \alpha $ and $L$, i.e.  $ q(z,\alpha,L) $ and $ \bar{H}(z,\alpha,L) $. In this work, we will not calculate the averaging length scale, $L$ but for the rest of our analysis we consider $L\sim 20$ Mpc which is estimated in  Ref. \cite{Atreya:2018iom}. Thus we find that the solutions for $ q$ and $ \bar{H}$ depend on only one free parameter, $ \alpha $, which will be estimated in next Section. 

\section{Analysis and Results}\label{sec:results}
In this Section, we will first estimate the value of the free parameter, $\alpha$ using the low redshift observations and standard $\Lambda$CDM model. Then, using the best fit value of $\alpha$, we will see the evolution of viscosity and bulk viscous EoS of the VSIDM fluid. 
\begin{figure}[]
	\includegraphics[width=0.6\linewidth]{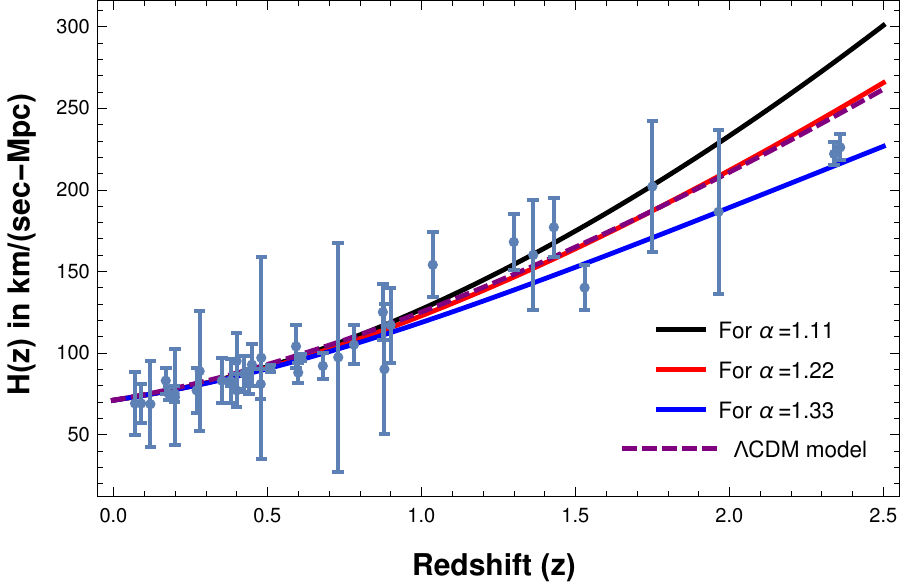}
	\caption{Hubble expansion rate, $H(z)$ obtained from our VSIDM model have been plotted with cosmic chronometer data and $\Lambda$CDM prediction as a function of redshift. Model with $\alpha=1.22$ fit the cosmic chronometer data and matches with the $\Lambda$CDM prediction. }
	\label{fig:h}
\end{figure}
\subsection{Hubble expansion rate}
\label{sus:hubexp}
In Fig. \ref{fig:h}, we have plotted the Hubble expansion rate, $H(z)$ as a function of redshift for different values of $\alpha$ with the cosmic chronometer data obtained from the Ref. \cite{Farooq:2016zwm}. The black, red and blue solid lines corresponds for $ \alpha=1.11, ~ \alpha=1.22 $ and $ \alpha=1.33 $, respectively.

To compare our viscous model with the standard cosmology, we have also plotted the Hubble expansion rate derived from the standard $\Lambda$CDM model (purple dashed line) in the Fig.\ref{fig:h}. Here we clearly see that the Hubble expansion depends on the dissipative strength of the dark matter, large is dissipation, larger will be the $ H(z) $. Although on small redshift, all models contribute equally to the Hubble rate, but on the large redshift due to difference into the dissipation term, all model contribute unequally in the $ H(z) $ and start deviating with each other.

We find that for $\alpha=1.11$ case, the $H(z)$ increases quickely and becomes very large at earlier times and does not fit the large redshift  Hubble data. Further, $ \alpha=1.33$ case the $H(z)$ increases slowly in comparison with the $\alpha=1.11$ case and does not fit the large redshift  Hubble data.  
But for $ \alpha=1.22$ case, the $H(z)$ explain the cosmic chronometer data and matches with the standard cosmological prediction.
\subsection{Fitting of Supernovae data}
In order to fit the supernovae data from VSIDM model, we calculate the measurable quantity distance modulus $\mu$. This is defined as  $\mu\equiv m-M$, where $m$ and $M$ represents the apparent and absolute magnitude of Type Ia supernovae (SNe Ia).
In terms of the luminosity distance $ d_{\mathrm{L}} $, the distance modulus, $\mu$ is defined as
\begin{equation}
\mu(z,\alpha)= 5\log_{10}{\left(\frac{\bar{d}_{\mathrm{L}}(z,\alpha)}{\mathrm{Mpc}} \right)} + 25~~,
\label{eq:distmod}
\end{equation}
where $ \bar{d}_{L}(z,\alpha)$ is a dimensionless luminosity distance, defined as $ \bar{d_{\mathrm{L}}}(z,\alpha) =  H_{0}~d_{\mathrm{L}} (z,\alpha)$. In order to calculate The luminosity distance is given as
\begin{equation}
d_{\mathrm{L}}(z,\alpha) =\frac{ (1+z)}{H_{0}}\int_{0}^{z}\frac{dz}{\bar{H}(z,\alpha)}
\label{eq:lumdist}
\end{equation}
\begin{figure}
	\includegraphics[width=0.6\linewidth]{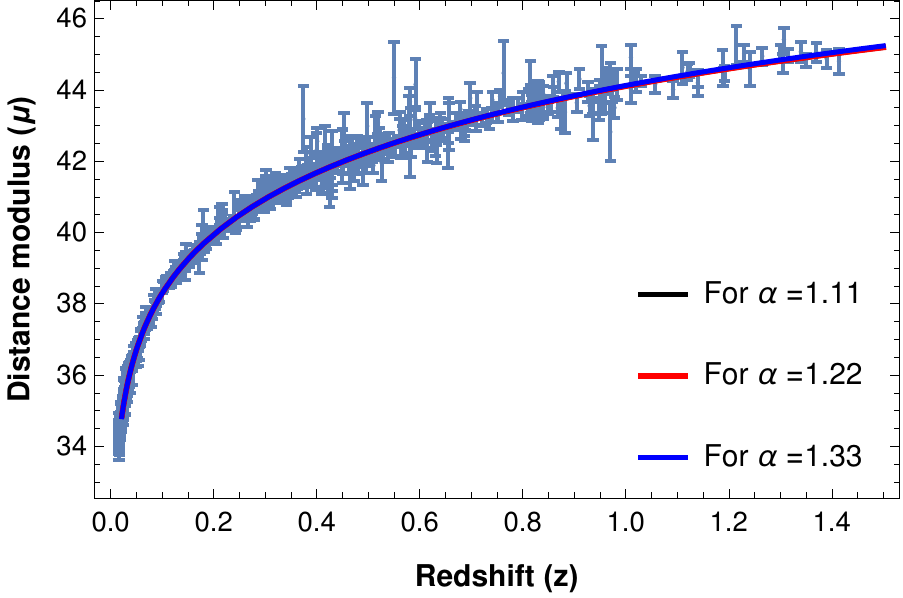}
\caption{Distance modulus, $\mu(z)$ obtained from VSIDM model have been plotted with the Union 2.1 SNe Ia data for different values of viscosity parameter, $\alpha$. The plot suggests that the VSIDM model fits the SNe Ia data very well. }
	\label{fig:dm}
\end{figure}
Further, using Eq. (\ref{eq:distmod}), we plot the distance modulus as a function of redshift for different values of $\alpha$ along with the SNe Ia data in Fig. \ref{fig:dm}. We take Union 2.1 compilation SNe Ia data from the Refs.\cite{Amanullah:2010vv,Suzuki:2011hu}, which consist of 580 SNe data. It  can be clearly seen that the values of $\alpha$ considered here, i.e. $\alpha=1.11, 1.22$ and $\alpha=1.33$ fit the SNe Ia data equally well. This implies that the fitting of SNe Ia data cannot suggest the correct evolution of the SIDM viscosity.
\subsection{Deceleration parameter $q$}
\label{sec:q}
\begin{figure}
	\includegraphics[width=0.6\linewidth]{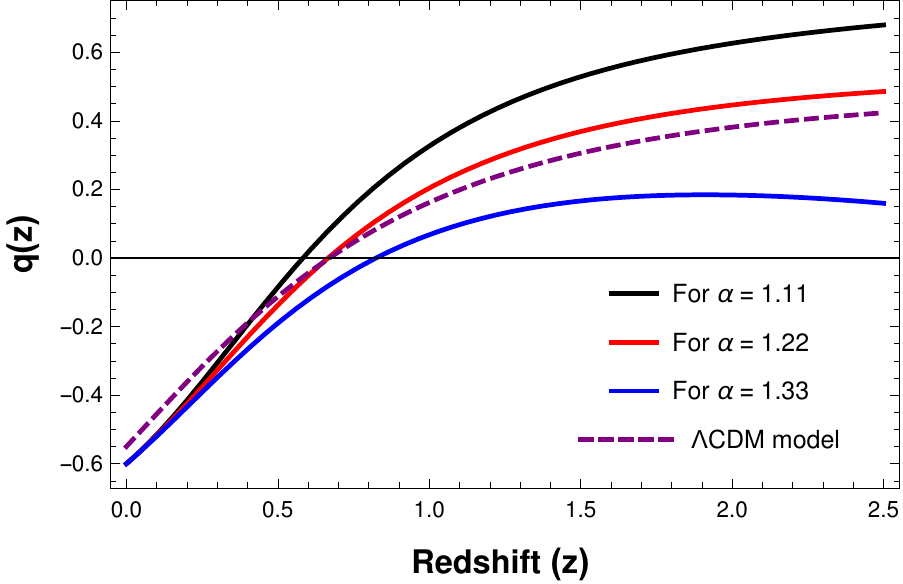}
	\caption{Deceleration parameter, $q(z)$ obtained from the VSIDM model parameters have been plotted with the $\Lambda$CDM model prediction.}
	\label{fig:q}
\end{figure}

The deceleration parameter ($q$) defined in the Eq. (\ref{eq:qev}) provides the information, whether the Universe is in the accelerating (for $q>0$) or in the decelerating phase (for $q<0$).
In order to see the epoch of decelerated to accelerated phase transition $z_{\mathrm{tr}}$ (i.e. epoch of $q>0$ to $q<0$) in our VSIDM model, we plot the $q(z)$ for different values of $\alpha$, as a function of redshift in Fig. \ref{fig:q}. The black, red and purple solid lines corresponds for $ \alpha=1.11, \alpha=1.22 $
and $ \alpha=1.33$, respectively. The green dashed line corresponds for $\Lambda$CDM prediction.
We can see that for large $\alpha$, the transition point is earlier (on large redshift). For $\alpha=1.11, ~1.22$ and $\alpha=1.33$, the transition points are $z_{\mathrm{tr}}=0.58, ~ z_{\mathrm{tr}}=0.66$ and $z_{\mathrm{tr}}= 0.81$, respectively. The transition point corresponding to $\alpha = 1.22$ matches with the $\Lambda$CDM model prediction.

We point out from Fig. \ref{fig:q} that for $\alpha=1.11$ case, the deceleration parameter increases and
settles around $q\sim 0.7$ for large redshift.  This is quite in contrast with
our expectation that $q$ should approach $0.5$ in the matter-dominated era. Also, it does not explain the large redshift Hubble data correctly, see \ref{sus:hubexp}. For $\alpha=1.33$ case, at higher redshift $q$ is decreasing and approaching towards $q\sim 0.16$ which is below the expected value of $q=0.5$ in matter dominated era. We may thus safely conclude that the case for $\alpha=1.11$ or $\alpha=1.33$ is surely not the case to appropriately describe the cosmic evolution.

Furthermore, from Fig. \ref{fig:q} we can see that for $\alpha=1.22$ case, the deceleration parameter saturates around $q\sim0.49$, which is very close to our expectation and slightly different from the $\Lambda$CDM model $q$ prediction. The important feature of this model is that the $H(z)$ obtained from this model overlaps with the $\Lambda$CDM expectation of Hubble parameter and explain the cosmic chronometer data correctly. Hence assuming $ C_{n} =0 $,  $\alpha=1.22$ is the most intriguing possibility to explain the Hubble data and $q$ value and matches with some of the $\Lambda$CDM prediction.  
\subsection{Statefinder technique}
As we have seen in the previous subsection that the best fit value of the model parameter for VSIDM matches with the $\Lambda$CDM model prediction. In order to see the deviation of the VSIDM model with the $\Lambda$CDM model, we adopt a geometric diagnostic approach as discussed in Ref. \cite{Sahni:2002fz}, which was introduced to differentiate between the different dark energy models. 
\begin{figure}[]
	\includegraphics[width=0.6\linewidth]{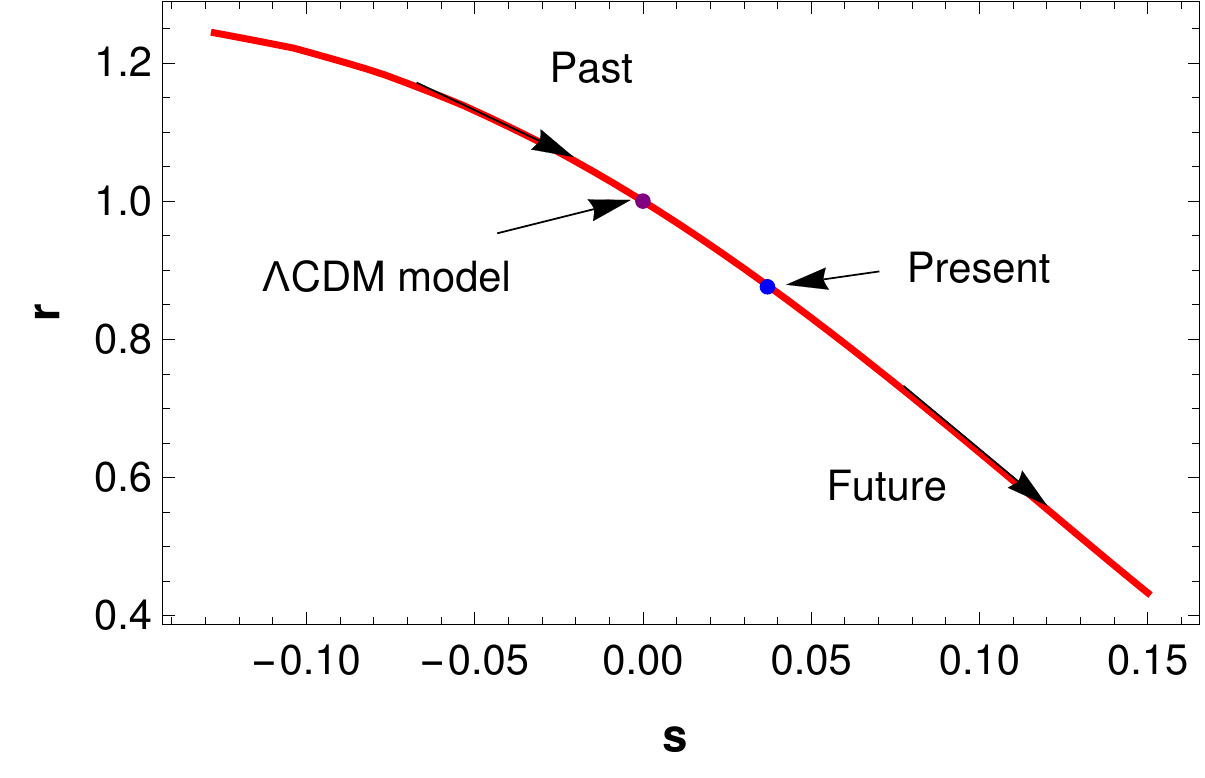}
	\caption{The Statefinder pair $(r,\mathrm{s}) $ evolution for the best fit model parameter $\alpha=1.22$ in the VSIDM model of the Universe.}  
	\label{fig:statef}
\end{figure}	 
In this approach, one calculates a Statefinder parameter pair $ \left\lbrace r, \mathrm{s} \right\rbrace $ which is related to the higher-order derivative of the Hubble expansion rate. 
In terms of the redshift, the Statefinder parameters are defined as
\begin{equation}
r(z) = 1 - 2(1+z)\left( \frac{H'}{H} \right) +(1+z)^{2}\left[\left( \frac{H'}{H}\right)^{2}+\frac{H''}{H} \right] 
\label{eq:rt}
\end{equation}
and
\begin{equation}
\mathrm{s}(z)=\frac{1}{3}\left( \frac{r(z)-1}{q(z)-\frac{1}{2}}\right) 
\label{eq:st}
\end{equation}
The idea lies in the fact that $ \left\lbrace r, s \right\rbrace $ pair is a fixed point given by $ \left\lbrace 1, 0 \right\rbrace $ for $\Lambda$CDM model , and may varies for the other models. In Figure \ref{fig:statef}, we plot the evolution of $r - \mathrm{s}$ plane for the best fit value of the parameter $\alpha$. We find that for our VSIDM model the pair lies on the second quadrant of $r - \mathrm{s}$ plane in the past and evolve towards the first quadrant. The present value of pair in VSIDM model $ \left\lbrace r, \mathrm{s} \right\rbrace $ is $ \left\lbrace 0.87,0.03 \right\rbrace $ which clearly implies that the VSIDM model is different from the $\Lambda$CDM model.
\subsection{Evolution of VSIDM viscosity on small redshift}
\begin{figure}[]
	\includegraphics[width=0.6\linewidth]{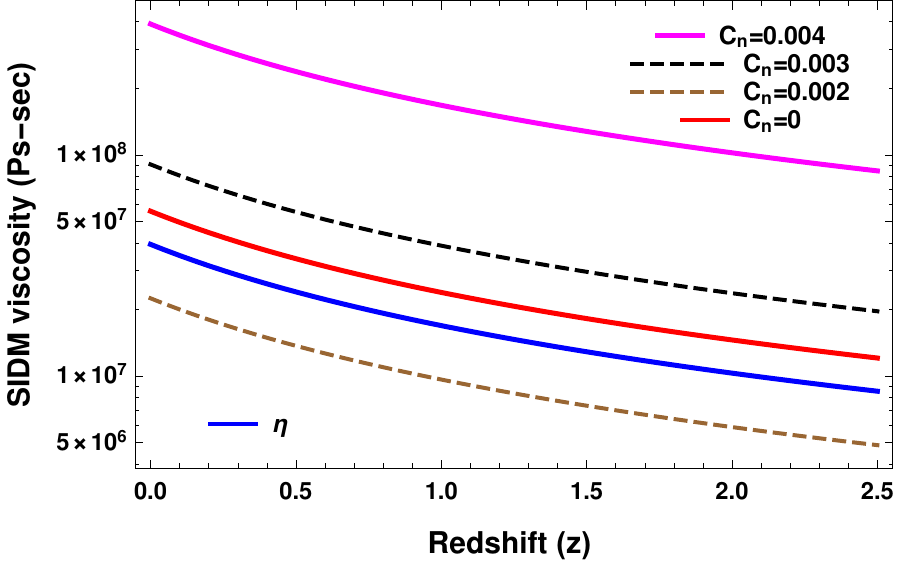}	
	\caption{The bulk viscosity ($\zeta$) and shear viscosity ($\eta$) of VSIDM have been plotted as a function of redshift for the best fit model parameter, $\alpha=1.22$. Blue line corresponds for shear viscosity and the rest of the lines corresponds for bulk viscosity at the different sound speeds.} 	
	\label{fig:bulkconst}
\end{figure}

To see the evolution of the bulk and shear viscosity of the VSIDM fluid for non vanishing sound speed, we fix $\frac{T}{m} = \frac{\langle v_{c}\rangle^{2}}{3} $ and assume redshift dependency of $ \frac{m}{\langle \sigma v \rangle} $  in equations (\ref{eq:zetas1}) and (\ref{eq:etas1}). In Fig. \ref{fig:bulkconst}, we plot, $ \eta $ and $ \zeta $ as a function of redshift for the best fit values of the viscosity parameter, i.e. $\alpha=1.22$. The blue line refer for the shear viscosity and the rest of the other lines corresponds for the bulk viscosity for different values of sound speed. We see that the $\zeta$ and $\eta$ are large at present, $ z=0 $ and decreases on the larger redshift $ z>0 $. This implies that at the earlier times when the halo starts forming the SIDM viscosities were small but at later times of cosmic evolution when the DM halo becomes more or less virialized, the viscosities contribute more. We also find that for small sound speed, $ C_{n} <0.027$, negative term in Eq. (\ref{eq:zetas1}) becomes large and decreases $\zeta$  in comparison with the $ C_{n} =0$ counterpart. But for large sound speed, $ C_{n} >0.027$, the positive term in Eq. (\ref {eq:zetas1}) becomes large and increases $\zeta$.  

Further, we emphasize that the value of the VSIDM viscosity obtained here can only be possible at the late times when the non-linear structure formation takes place and collapse objects are formed. Otherwise, as shown in Refs. \cite{Velten:2013pra, Velten:2014xca} for sufficiently large DM viscosity at earlier times, the density perturbation may wash out and non-linear structure formation will not be possible.

As we have discussed above, at the present ($z=0$), the viscous contribution from the bulk as well as shear DM viscosity increases on low redshift thus we may expect some consequences. In the Ref. \cite{Atreya:2017pny}, we have shown that at the present time, $ z=0 $ the viscous effects of VSIDM are large and can explain the present observed acceleration. Further, these results also provide us the physical basis of the cosmic acceleration and also why it starts at a late time (low redshift), not at an early time (large redshift).
\subsection{Bulk viscosity EoS}
\label{subsec:EoS}
\begin{figure}%
	\centering
	\subfigure[]{
		\label{fig:EoS}
		\includegraphics[height=2.05in,width=3.1in]{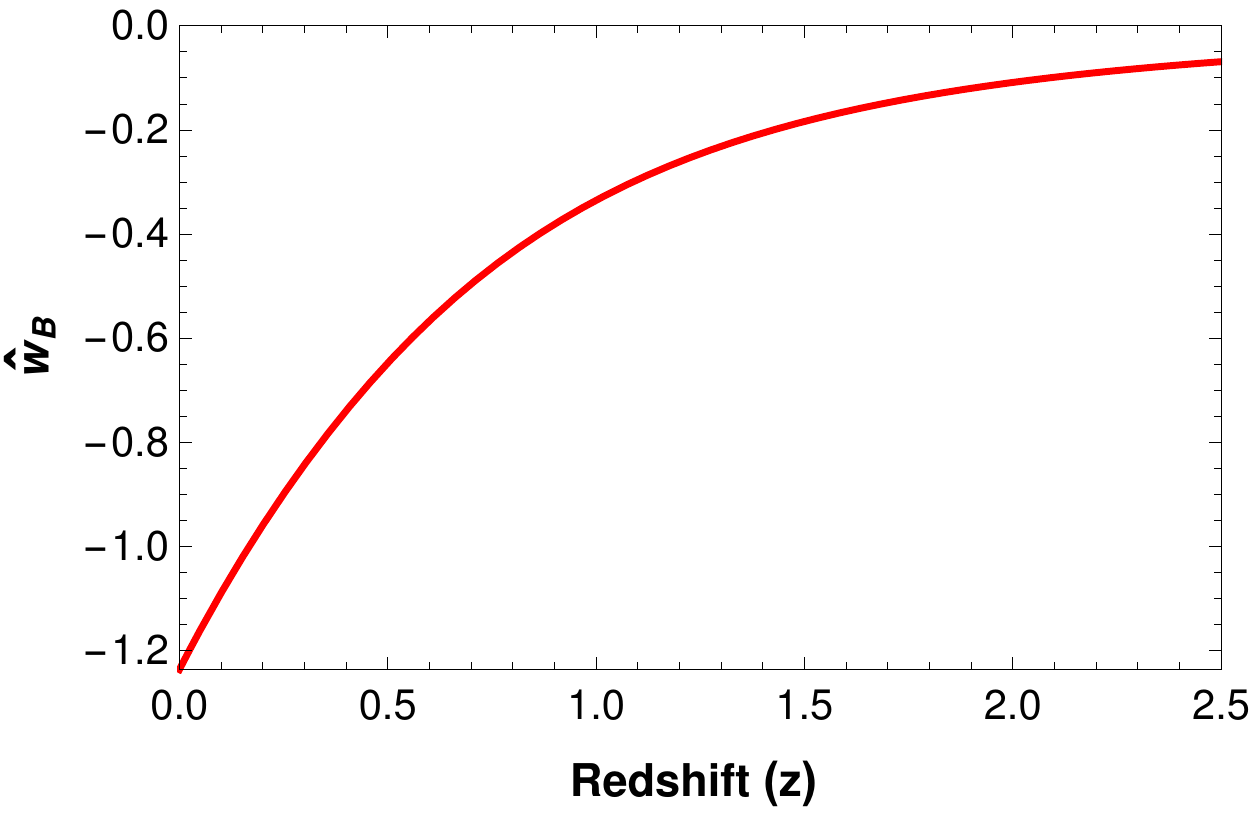}}
	\subfigure[]{
		\vspace{-0.3cm}	
		\label{fig:age}	\includegraphics[height=2in,width=3.1in]{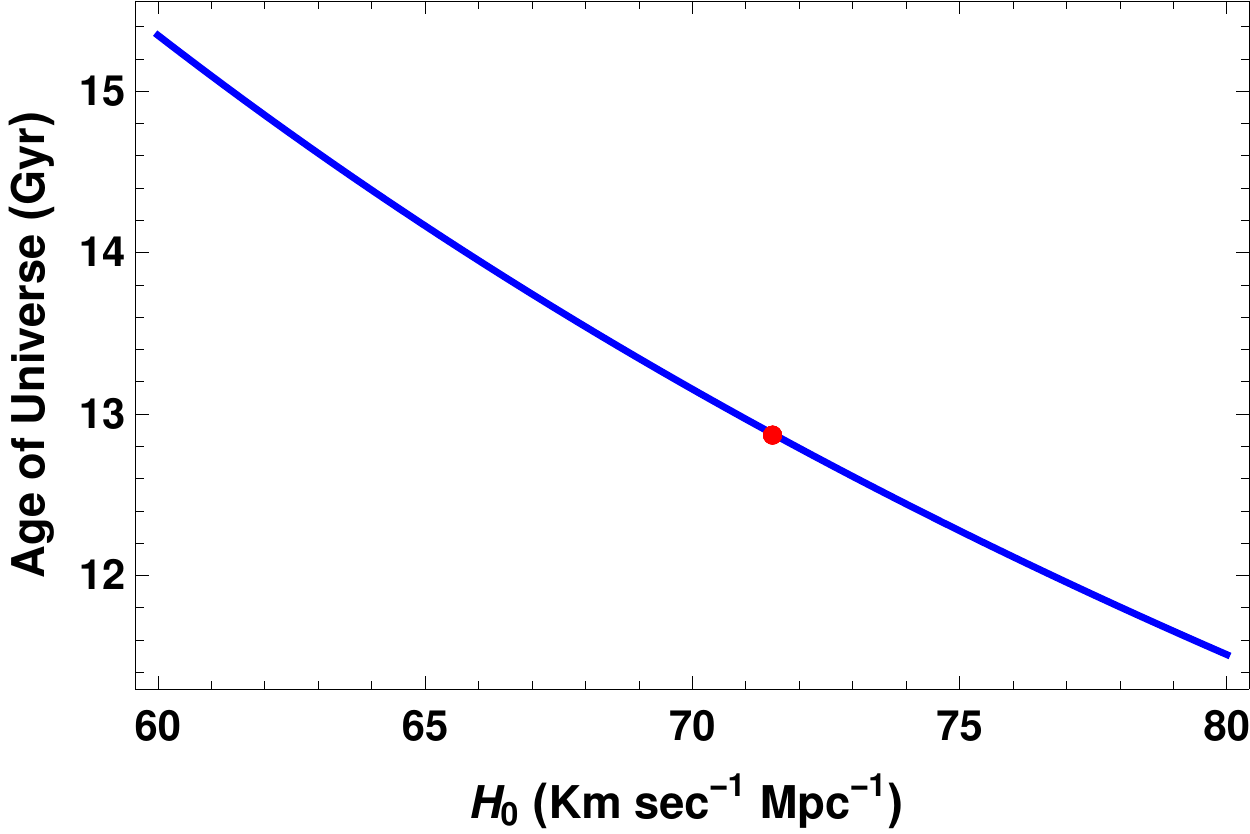}} \\
	\caption{In Fig. \ref{fig:EoS} equation of state, $ \hat{w}_{B}$ of VSIDM model as a function of redshift for best fit model parameter. In Fig. \ref{fig:age}, the age of the Universe is plotted as a function of the Hubble expansion rate. The red point corresponds for $ t_{\mathrm{U}}=13 $ Gyr and  $ H_{0} =71.5$ Km sec$^{-1}$Mpc$^{-1}$ obtained from the best fit model parameter.} 
\end{figure}

In this subsection, we show the evolution of EoS of VSIDM fluid on small redshift. In Fig. \ref{fig:EoS}, we plot the equation of state corresponding the SIDM bulk viscosity,  $ \hat{w}_{B}$ as a function of the redshift for the best fit value of viscosity parameter, i.e. $ \alpha = 1.22$. We see that on the small redshift $ \hat{w}_{B}$ subsequently becomes more negative and at present,  $ \hat{w}_{B}(z=0)=-1.2$  and  on large redshift, $ \hat{w}_{B}$ increases and approaches towards $ \hat{w}_{B}\sim 0$. 
\subsection{Age of Universe}
The age of Universe at any redshift, $ t_{\mathrm{U}}(z) $ is obtained from the Hubble expansion rate as 
\begin{equation}
t_{\mathrm{U}}(z)=\int_{z}^{\infty}~\frac{dz}{(1+z)H(z)}
\end{equation} 
In this work, we have assumed that the SIDM viscosity becomes effective only at late time $ z\leq 2.5 $ and consequently, the viscous effect modifies the evolution of the Universe only at a late time. At early time $  z>2.5 $, the evolution of the Universe is governed  via standard cosmology. Thus we consider
\begin{equation}
H(z) = \left\{ \,
\begin{IEEEeqnarraybox}[][c]{l?s}
\IEEEstrut H_{\mathrm{visc}} & if $ z\leq2.5 $, \\H_{\mathrm{\Lambda CDM}} & if $z>2.5$, \IEEEstrut
\end{IEEEeqnarraybox}\right.
\label{eq:Hubble}
\end{equation}
where $ H_{\mathrm{visc}} $ is obtained from the Eqs. (\ref{eq:qev}) and (\ref{eq:hevz}) and $ H_{\mathrm{\Lambda CDM}}\approx H_{0}\big[ \Omega_{B}(z) + \Omega_{\chi}(z) \big]^{1/2}$. Thus, using the best fit value of $ \alpha $, we get
\begin{equation}
t_{\mathrm{U}}\sim \frac{ 0.974}{H_{0}}. 
\end{equation}

In Fig. \ref{fig:age}, we plot the age of the Univere,  $ t_{\mathrm{U}} $  in the VSIDM model as a function of the Hubble expansion rate. We see that as the $ H_{0} $ increases, $ t_{\mathrm{U}} $ decreases. In VSIDM model, using the best fit value of model parameter,  $ \alpha=1.22 $, $t_{\mathrm{U}}=13 $ Gyr. Our estimation of  $t_{\mathrm{U}}$ is slightly small in comparison with the age of the Univere ($13.76 $ Gyr) obtained in the CMB anisotropy data \cite{Tegmark:2006az} and larger than the age of globular cluster ($12.9 $ Gyr) \cite{Carretta:1999ii}.

\section{Conclusion}
\label{sec:danc}
The self-interacting dark matter  may solve the small scale problems of collision-less cold dark matter. In the case if SIDM fluid is viscous, it may affect the cosmic evolution history and explain the present observed accelerated expansion of the Universe. In this work, we study the evolution of the viscous effect of SIDM from the low reshift observational data.

We calculate the bulk viscosity for non-vanishing sound speed of dark matter using the kinetic theory and relaxation time approximation. The VSIDM viscosities are calculated in the non-linear regime, assuming that the DM halos are gravitationally bound and may not be completely virialized. We check the dependency of bulk viscosity on sound speed and show that for small sound speed, $ C_{n}<0.0027 $, $ \zeta $ is small but for large sound speed $ C_{n}>0.0027 $, $ \zeta $ becomes large in comparison with the $ C_{n}=0 $ case. Then, using the KSS bound, $\eta/\mathfrak{s}\geq \frac{1}{4\pi}$, we derive a new constraint on $ \sigma/m $, as $ \frac{\sigma }{m}\leq \frac{(2\pi)^{\frac{5}{2}}}{(3)^{\frac{1}{2}}}\left( \frac{1}{m}\right)^{3} $. Further, using the value of  $ \sigma/m $ obtained from the numerical simulations, we show that KSS bound allowed a sub-GeV ($ \mathcal{O} (0.1)$ GeV) mass of the SIDM particle. In future, this limit will further be improved for more precise estimation of $\sigma/m $. 

To study the evolution of SIDM viscosity at the redshift of our interest, we assume the power law form of bulk $ \zeta(z) =\zeta_{0}\left(a/a_{0}\right) ^{\alpha} $ and shear viscosity $ \eta(z) =\eta_{0}\left(a/a_{0}\right) ^{\alpha} $. Then inspired from the observational evidence that velocity gradient is constant on typical cluster and supercluster scale at the present time, we assume it to be constant on the low redshift interval $ 0 \le z \le 2.5 $.

Further, we calculate the Hubble expansion rate and deceleration parameter, which depends on the viscosity parameter, $\alpha$ and the length scale, $L$. We assume $L=20$ Mpc which is larger than the typical cluster size DM halo. Further, using the cosmic chronometer data points and the correct value of the deceleration parameter at the matter-dominated era, we obtain $ \alpha = 1.22$. The best fit values of model parameter shows that the viscous coefficients, $ \eta $ and $ \zeta $ are large at present, $ z=0 $ and decrease at earlier time $ z>0 $. The deceleration to acceleration transition point in this model is $z_{\mathrm{tr}}= 0.66$, which matches with the $\Lambda$CDM model. We find that the VSIDM model fits with the supernovae data very well. Although our VSIDM model matches the $ \Lambda$CDM prediction, but using the Statefinder technique we find that our model is different from the $ \Lambda$CDM model. Also in our model, the age of the Universe is  $13 $ Gyr, which is smaller than the age inferred from the CMB anisotropy data but larger than the globular cluster age. 

Thus we conclude the VSIDM model can unify the dark sectors (DM and dark energy) and may be a possible alternative theory of the standard model of cosmology at a small redshift. Our result can also provide a new DM mass range, which will be crucial for future particle dark matter searches. 

\section{Acknowledgements}
I would like to thank Prof. Jitesh R. Bhatt, Prof. Subhendra Mohanty and Dr. Abhishek Atreya for useful discussions and comments. I would also like to thank Richa Arya for reading the manuscript and providing useful suggestions.
 
 \bibliographystyle{utphys}
 \bibliography{viscevolution}

\providecommand{\href}[2]{#2}\begingroup\raggedright\begin{thebibliography}{10}

\bibitem{Weinberg:2013aya}
D.~H. Weinberg, J.~S. Bullock, F.~Governato, R.~Kuzio~de Naray, and A.~H.~G.
  Peter, ``{Cold dark matter: controversies on small scales},''
  \href{http://dx.doi.org/10.1073/pnas.1308716112}{{\em Proc. Nat. Acad. Sci.}
  {\bfseries 112} (2015) 12249--12255},
\href{http://arxiv.org/abs/1306.0913}{{\ttfamily arXiv:1306.0913
  [astro-ph.CO]}}.
%%CITATION = ARXIV:1306.0913;%%.

\bibitem{Tulin:2017ara}
S.~Tulin and H.-B. Yu, ``{Dark Matter Self-interactions and Small Scale
  Structure},'' \href{http://dx.doi.org/10.1016/j.physrep.2017.11.004}{{\em
  Phys. Rept.} {\bfseries 730} (2018) 1--57},
\href{http://arxiv.org/abs/1705.02358}{{\ttfamily arXiv:1705.02358 [hep-ph]}}.
%%CITATION = ARXIV:1705.02358;%%.

\bibitem{Spergel:1999mh}
D.~N. Spergel and P.~J. Steinhardt, ``{Observational evidence for
  selfinteracting cold dark matter},''
  \href{http://dx.doi.org/10.1103/PhysRevLett.84.3760}{{\em Phys. Rev. Lett.}
  {\bfseries 84} (2000) 3760--3763},
\href{http://arxiv.org/abs/astro-ph/9909386}{{\ttfamily arXiv:astro-ph/9909386
  [astro-ph]}}.
%%CITATION = ASTRO-PH/9909386;%%.

\bibitem{Wandelt:2000ad}
B.~D. Wandelt, R.~Dave, G.~R. Farrar, P.~C. McGuire, D.~N. Spergel, and P.~J.
  Steinhardt, ``{Selfinteracting dark matter},'' in {\em {Sources and detection
  of dark matter and dark energy in the universe. Proceedings, 4th
  International Symposium, DM 2000, Marina del Rey, USA, February 23-25,
  2000}}, pp.~263--274.
\newblock 2000.
\newblock \href{http://arxiv.org/abs/astro-ph/0006344}{{\ttfamily
  arXiv:astro-ph/0006344 [astro-ph]}}.
\newblock
\url{http://www.slac.stanford.edu/spires/find/books/www?cl=QB461:I57:2000}.
\newblock
%%CITATION = ASTRO-PH/0006344;%%.

\bibitem{Randall:2007ph}
S.~W. Randall, M.~Markevitch, D.~Clowe, A.~H. Gonzalez, and M.~Bradac,
  ``{Constraints on the Self-Interaction Cross-Section of Dark Matter from
  Numerical Simulations of the Merging Galaxy Cluster 1E 0657-56},''
  \href{http://dx.doi.org/10.1086/587859}{{\em Astrophys. J.} {\bfseries 679}
  (2008) 1173--1180},
\href{http://arxiv.org/abs/0704.0261}{{\ttfamily arXiv:0704.0261 [astro-ph]}}.
%%CITATION = ARXIV:0704.0261;%%.

\bibitem{Tulin:2013teo}
S.~Tulin, H.-B. Yu, and K.~M. Zurek, ``{Beyond Collisionless Dark Matter:
  Particle Physics Dynamics for Dark Matter Halo Structure},''
  \href{http://dx.doi.org/10.1103/PhysRevD.87.115007}{{\em Phys. Rev.}
  {\bfseries D87} no.~11, (2013) 115007},
\href{http://arxiv.org/abs/1302.3898}{{\ttfamily arXiv:1302.3898 [hep-ph]}}.
%%CITATION = ARXIV:1302.3898;%%.

\bibitem{Sarkar2018}
S.~Sarkar, ``Is dark matter self-interacting?,''
  \href{http://dx.doi.org/10.1038/s41550-018-0598-6}{{\em Nature Astronomy}
  {\bfseries 2} no.~11, (Oct., 2018) 856--857}.
  \url{https://doi.org/10.1038/s41550-018-0598-6}.

\bibitem{Atreya:2017pny}
A.~Atreya, J.~R. Bhatt, and A.~Mishra, ``{Viscous Self Interacting Dark Matter
  and Cosmic Acceleration},''
  \href{http://dx.doi.org/10.1088/1475-7516/2018/02/024}{{\em JCAP} {\bfseries
  1802} no.~02, (2018) 024},
\href{http://arxiv.org/abs/1709.02163}{{\ttfamily arXiv:1709.02163
  [astro-ph.CO]}}.
%%CITATION = ARXIV:1709.02163;%%.

\bibitem{Atreya:2018iom}
A.~Atreya, J.~R. Bhatt, and A.~K. Mishra, ``{Viscous Self Interacting Dark
  Matter Cosmology For Small Redshift},''
  \href{http://dx.doi.org/10.1088/1475-7516/2019/02/045}{{\em JCAP} {\bfseries
  1902} (2019) 045},
\href{http://arxiv.org/abs/1810.11666}{{\ttfamily arXiv:1810.11666
  [astro-ph.CO]}}.
%%CITATION = ARXIV:1810.11666;%%.

\bibitem{Padmanabhan:1987dg}
T.~Padmanabhan and S.~M. Chitre, ``{Viscous universes},''
\href{http://dx.doi.org/10.1016/0375-9601(87)90104-6}{{\em Phys. Lett.}
  {\bfseries A120} (1987) 433--436}.
%%CITATION = PHLTA,A120,433;%%.

\bibitem{Gron:1990ew}
O.~Gron, ``{Viscous inflationary universe models},''
\href{http://dx.doi.org/10.1007/BF00643930}{{\em Astrophys. Space Sci.}
  {\bfseries 173} (1990) 191--225}.
%%CITATION = APSSB,173,191;%%.

\bibitem{Cheng:1991uu}
B.~Cheng, ``{Bulk viscosity in the early universe},''
\href{http://dx.doi.org/10.1016/0375-9601(91)90660-Z}{{\em Phys. Lett.}
  {\bfseries A160} (1991) 329--338}.
%%CITATION = PHLTA,A160,329;%%.

\bibitem{Zimdahl:1996ka}
W.~Zimdahl, ``{Bulk viscous cosmology},''
  \href{http://dx.doi.org/10.1103/PhysRevD.53.5483}{{\em Phys. Rev.} {\bfseries
  D53} (1996) 5483--5493},
\href{http://arxiv.org/abs/astro-ph/9601189}{{\ttfamily arXiv:astro-ph/9601189
  [astro-ph]}}.
%%CITATION = ASTRO-PH/9601189;%%.

\bibitem{Fabris:2005ts}
J.~C. Fabris, S.~V.~B. Goncalves, and R.~de~Sa~Ribeiro, ``{Bulk viscosity
  driving the acceleration of the Universe},''
  \href{http://dx.doi.org/10.1007/s10714-006-0236-y}{{\em Gen. Rel. Grav.}
  {\bfseries 38} (2006) 495--506},
\href{http://arxiv.org/abs/astro-ph/0503362}{{\ttfamily arXiv:astro-ph/0503362
  [astro-ph]}}.
%%CITATION = ASTRO-PH/0503362;%%.

\bibitem{Mathews:2008hk}
G.~J. Mathews, N.~Q. Lan, and C.~Kolda, ``{Late Decaying Dark Matter, Bulk
  Viscosity and the Cosmic Acceleration},''
  \href{http://dx.doi.org/10.1103/PhysRevD.78.043525}{{\em Phys. Rev.}
  {\bfseries D78} (2008) 043525},
\href{http://arxiv.org/abs/0801.0853}{{\ttfamily arXiv:0801.0853 [astro-ph]}}.
%%CITATION = ARXIV:0801.0853;%%.

\bibitem{Avelino:2008ph}
A.~Avelino and U.~Nucamendi, ``{Can a matter-dominated model with constant bulk
  viscosity drive the accelerated expansion of the universe?},''
  \href{http://dx.doi.org/10.1088/1475-7516/2009/04/006}{{\em JCAP} {\bfseries
  0904} (2009) 006},
\href{http://arxiv.org/abs/0811.3253}{{\ttfamily arXiv:0811.3253 [gr-qc]}}.
%%CITATION = ARXIV:0811.3253;%%.

\bibitem{Das:2008mj}
S.~Das and N.~Banerjee, ``{Can neutrino viscosity drive the late time cosmic
  acceleration?},'' \href{http://dx.doi.org/10.1007/s10773-012-1152-4}{{\em
  Int. J. Theor. Phys.} {\bfseries 51} (2012) 2771--2778},
\href{http://arxiv.org/abs/0806.3666}{{\ttfamily arXiv:0806.3666 [gr-qc]}}.
%%CITATION = ARXIV:0806.3666;%%.

\bibitem{Piattella:2011bs}
O.~F. Piattella, J.~C. Fabris, and W.~Zimdahl, ``{Bulk viscous cosmology with
  causal transport theory},''
  \href{http://dx.doi.org/10.1088/1475-7516/2011/05/029}{{\em JCAP} {\bfseries
  1105} (2011) 029},
\href{http://arxiv.org/abs/1103.1328}{{\ttfamily arXiv:1103.1328
  [astro-ph.CO]}}.
%%CITATION = ARXIV:1103.1328;%%.

\bibitem{Velten:2011bg}
H.~Velten and D.~J. Schwarz, ``{Constraints on dissipative unified dark
  matter},'' \href{http://dx.doi.org/10.1088/1475-7516/2011/09/016}{{\em JCAP}
  {\bfseries 1109} (2011) 016},
\href{http://arxiv.org/abs/1107.1143}{{\ttfamily arXiv:1107.1143
  [astro-ph.CO]}}.
%%CITATION = ARXIV:1107.1143;%%.

\bibitem{Gagnon:2011id}
J.-S. Gagnon and J.~Lesgourgues, ``{Dark goo: Bulk viscosity as an alternative
  to dark energy},''
  \href{http://dx.doi.org/10.1088/1475-7516/2011/09/026}{{\em JCAP} {\bfseries
  1109} (2011) 026},
\href{http://arxiv.org/abs/1107.1503}{{\ttfamily arXiv:1107.1503
  [astro-ph.CO]}}.
%%CITATION = ARXIV:1107.1503;%%.

\bibitem{Mohan:2017poq}
N.~D.~J. Mohan, A.~Sasidharan, and T.~K. Mathew, ``{Bulk viscous matter and
  recent acceleration of the universe based on causal viscous theory},''
  \href{http://dx.doi.org/10.1140/epjc/s10052-017-5428-y}{{\em Eur. Phys. J.}
  {\bfseries C77} no.~12, (2017) 849},
\href{http://arxiv.org/abs/1708.02437}{{\ttfamily arXiv:1708.02437 [gr-qc]}}.
%%CITATION = ARXIV:1708.02437;%%.

\bibitem{Cruz:2018yrr}
N.~Cruz, E.~González, S.~Lepe, and D.~Sáez-Chillón~Gómez, ``{Analysing
  dissipative effects in the $\Lambda$CDM model},''
  \href{http://dx.doi.org/10.1088/1475-7516/2018/12/017}{{\em JCAP} {\bfseries
  1812} no.~12, (2018) 017},
\href{http://arxiv.org/abs/1807.10729}{{\ttfamily arXiv:1807.10729 [gr-qc]}}.
%%CITATION = ARXIV:1807.10729;%%.

\bibitem{Li:2009mf}
B.~Li and J.~D. Barrow, ``{Does Bulk Viscosity Create a Viable Unified Dark
  Matter Model?},'' \href{http://dx.doi.org/10.1103/PhysRevD.79.103521}{{\em
  Phys. Rev.} {\bfseries D79} (2009) 103521},
\href{http://arxiv.org/abs/0902.3163}{{\ttfamily arXiv:0902.3163 [gr-qc]}}.
%%CITATION = ARXIV:0902.3163;%%.

\bibitem{Barbosa:2015ndx}
C.~M.~S. Barbosa, J.~C. Fabris, O.~F. Piattella, H.~E.~S. Velten, and
  W.~Zimdahl, ``{Viscous Cosmology},'' in {\em {Proceedings, 12th International
  Conference on Gravitation, Astrophysics and Cosmology (ICGAC-12): Moscow,
  Russia, June 28-July 5, 2015}}.
\newblock 2015.
\newblock
\href{http://arxiv.org/abs/1512.00921}{{\ttfamily arXiv:1512.00921
  [astro-ph.CO]}}.
\newblock
%%CITATION = ARXIV:1512.00921;%%.

\bibitem{Floerchinger:2014jsa}
S.~Floerchinger, N.~Tetradis, and U.~A. Wiedemann, ``{Accelerating Cosmological
  Expansion from Shear and Bulk Viscosity},''
  \href{http://dx.doi.org/10.1103/PhysRevLett.114.091301}{{\em Phys. Rev.
  Lett.} {\bfseries 114} no.~9, (2015) 091301},
\href{http://arxiv.org/abs/1411.3280}{{\ttfamily arXiv:1411.3280 [gr-qc]}}.
%%CITATION = ARXIV:1411.3280;%%.

\bibitem{Rezaei:2019ruy}
Z.~Rezaei, ``{Accelerated expansion of the Universe in the Presence of Dark
  Matter Pressure},''
\href{http://arxiv.org/abs/1906.08648}{{\ttfamily arXiv:1906.08648 [gr-qc]}}.
%%CITATION = ARXIV:1906.08648;%%.

\bibitem{Anand:2017wsj}
S.~Anand, P.~Chaubal, A.~Mazumdar, and S.~Mohanty, ``{Cosmic viscosity as a
  remedy for tension between PLANCK and LSS data},''
  \href{http://dx.doi.org/10.1088/1475-7516/2017/11/005}{{\em JCAP} {\bfseries
  1711} no.~11, (2017) 005},
\href{http://arxiv.org/abs/1708.07030}{{\ttfamily arXiv:1708.07030
  [astro-ph.CO]}}.
%%CITATION = ARXIV:1708.07030;%%.

\bibitem{Bhatt:2019qbq}
J.~R. Bhatt, A.~K. Mishra, and A.~C. Nayak, ``{Viscous dark matter and 21 cm
  cosmology},''
\href{http://arxiv.org/abs/1901.08451}{{\ttfamily arXiv:1901.08451
  [astro-ph.CO]}}.
%%CITATION = ARXIV:1901.08451;%%.

\bibitem{Mishra:2019uxl}
A.~K. Mishra, ``{Lightening the Dark Matter from its Viscosity and Explanation
  of EDGES Anomaly},''
\href{http://arxiv.org/abs/1907.04238}{{\ttfamily arXiv:1907.04238
  [astro-ph.CO]}}.
%%CITATION = ARXIV:1907.04238;%%.

\bibitem{Velten:2013pra}
H.~Velten, D.~J. Schwarz, J.~C. Fabris, and W.~Zimdahl, ``{Viscous dark matter
  growth in (neo-)Newtonian cosmology},''
  \href{http://dx.doi.org/10.1103/PhysRevD.88.103522}{{\em Phys. Rev.}
  {\bfseries D88} no.~10, (2013) 103522},
\href{http://arxiv.org/abs/1307.6536}{{\ttfamily arXiv:1307.6536
  [astro-ph.CO]}}.
%%CITATION = ARXIV:1307.6536;%%.

\bibitem{Velten:2014xca}
H.~Velten, T.~R.~P. Caramês, J.~C. Fabris, L.~Casarini, and R.~C. Batista,
  ``{Structure formation in a $\Lambda$ viscous CDM universe},''
  \href{http://dx.doi.org/10.1103/PhysRevD.90.123526}{{\em Phys. Rev.}
  {\bfseries D90} no.~12, (2014) 123526},
\href{http://arxiv.org/abs/1410.3066}{{\ttfamily arXiv:1410.3066
  [astro-ph.CO]}}.
%%CITATION = ARXIV:1410.3066;%%.

\bibitem{Goswami:2016tsu}
G.~Goswami, G.~K. Chakravarty, S.~Mohanty, and A.~R. Prasanna, ``{Constraints
  on cosmological viscosity and self interacting dark matter from gravitational
  wave observations},''
  \href{http://dx.doi.org/10.1103/PhysRevD.95.103509}{{\em Phys. Rev.}
  {\bfseries D95} no.~10, (2017) 103509},
\href{http://arxiv.org/abs/1603.02635}{{\ttfamily arXiv:1603.02635 [hep-ph]}}.
%%CITATION = ARXIV:1603.02635;%%.

\bibitem{Lu:2018smr}
B.-Q. Lu, D.~Huang, Y.-L. Wu, and Y.-F. Zhou, ``{Damping of gravitational waves
  in a viscous Universe and its implication for dark matter
  self-interactions},''
\href{http://arxiv.org/abs/1803.11397}{{\ttfamily arXiv:1803.11397
  [astro-ph.HE]}}.
%%CITATION = ARXIV:1803.11397;%%.

\bibitem{Brevik:2019yma}
I.~Brevik and S.~Nojiri, ``{Gravitational Waves in the Presence of
  Viscosity},'' \href{http://dx.doi.org/10.1142/S0218271819501335}{{\em Int. J.
  Mod. Phys.} {\bfseries D28} no.~10, (2019) 1950133},
\href{http://arxiv.org/abs/1901.00767}{{\ttfamily arXiv:1901.00767 [gr-qc]}}.
%%CITATION = ARXIV:1901.00767;%%.

\bibitem{Cai:2017buj}
R.-G. Cai, T.-B. Liu, and S.-J. Wang, ``{Gravitational wave as probe of
  superfluid dark matter},''
  \href{http://dx.doi.org/10.1103/PhysRevD.97.023027}{{\em Phys. Rev.}
  {\bfseries D97} no.~2, (2018) 023027},
\href{http://arxiv.org/abs/1710.02425}{{\ttfamily arXiv:1710.02425 [hep-ph]}}.
%%CITATION = ARXIV:1710.02425;%%.

\bibitem{Anand:2017ktp}
S.~Anand, P.~Chaubal, A.~Mazumdar, S.~Mohanty, and P.~Parashari, ``{Bounds on
  Neutrino Mass in Viscous Cosmology},''
  \href{http://dx.doi.org/10.1088/1475-7516/2018/05/031}{{\em JCAP} {\bfseries
  1805} no.~05, (2018) 031},
\href{http://arxiv.org/abs/1712.01254}{{\ttfamily arXiv:1712.01254
  [astro-ph.CO]}}.
%%CITATION = ARXIV:1712.01254;%%.

\bibitem{Medina:2019skp}
S.~B. Medina, M.~Nowakowski, and D.~Batic, ``{Viscous Cosmologies},''
\href{http://arxiv.org/abs/1901.09787}{{\ttfamily arXiv:1901.09787 [gr-qc]}}.
%%CITATION = ARXIV:1901.09787;%%.

\bibitem{Yang:2019qza}
W.~Yang, S.~Pan, E.~Di~Valentino, A.~Paliathanasis, and J.~Lu, ``{Challenging
  bulk viscous unified scenarios with cosmological observations},''
\href{http://arxiv.org/abs/1906.04162}{{\ttfamily arXiv:1906.04162
  [astro-ph.CO]}}.
%%CITATION = ARXIV:1906.04162;%%.

\bibitem{Bhatt:2019yld}
J.~R. Bhatt, P.~K. Natwariya, and A.~K. Pandey, ``{Viscosity in cosmic
  fluids},''
\href{http://arxiv.org/abs/1907.03445}{{\ttfamily arXiv:1907.03445
  [astro-ph.CO]}}.
%%CITATION = ARXIV:1907.03445;%%.

\bibitem{Brevik:2017msy}
I.~Brevik, Ã.~Grøn, J.~de~Haro, S.~D. Odintsov, and E.~N. Saridakis,
  ``{Viscous Cosmology for Early- and Late-Time Universe},''
  \href{http://dx.doi.org/10.1142/S0218271817300245}{{\em Int. J. Mod. Phys.}
  {\bfseries D26} no.~14, (2017) 1730024},
\href{http://arxiv.org/abs/1706.02543}{{\ttfamily arXiv:1706.02543 [gr-qc]}}.
%%CITATION = ARXIV:1706.02543;%%.

\bibitem{Kovtun:2004de}
P.~Kovtun, D.~T. Son, and A.~O. Starinets, ``{Viscosity in strongly interacting
  quantum field theories from black hole physics},''
  \href{http://dx.doi.org/10.1103/PhysRevLett.94.111601}{{\em Phys. Rev. Lett.}
  {\bfseries 94} (2005) 111601},
\href{http://arxiv.org/abs/hep-th/0405231}{{\ttfamily arXiv:hep-th/0405231
  [hep-th]}}.
%%CITATION = HEP-TH/0405231;%%.

\bibitem{Tegmark:2006az}
{\bfseries SDSS} Collaboration, M.~Tegmark {\em et~al.}, ``{Cosmological
  Constraints from the SDSS Luminous Red Galaxies},''
  \href{http://dx.doi.org/10.1103/PhysRevD.74.123507}{{\em Phys. Rev.}
  {\bfseries D74} (2006) 123507},
\href{http://arxiv.org/abs/astro-ph/0608632}{{\ttfamily arXiv:astro-ph/0608632
  [astro-ph]}}.
%%CITATION = ASTRO-PH/0608632;%%.

\bibitem{Carretta:1999ii}
E.~Carretta, R.~G. Gratton, G.~Clementini, and F.~Fusi~Pecci, ``{Distances,
  ages and epoch of formation of globular clusters},''
  \href{http://dx.doi.org/10.1086/308629}{{\em Astrophys. J.} {\bfseries 533}
  (2000) 215--235},
\href{http://arxiv.org/abs/astro-ph/9902086}{{\ttfamily arXiv:astro-ph/9902086
  [astro-ph]}}.
%%CITATION = ASTRO-PH/9902086;%%.

\bibitem{Ahn:2004xt}
K.-J. Ahn and P.~R. Shapiro, ``{Formation and evolution of the self-interacting
  dark matter halos},''
  \href{http://dx.doi.org/10.1111/j.1365-2966.2005.09492.x}{{\em Mon. Not. Roy.
  Astron. Soc.} {\bfseries 363} (2005) 1092--1124},
\href{http://arxiv.org/abs/astro-ph/0412169}{{\ttfamily arXiv:astro-ph/0412169
  [astro-ph]}}.
%%CITATION = ASTRO-PH/0412169;%%.

\bibitem{Rocha:2012jg}
M.~Rocha, A.~H.~G. Peter, J.~S. Bullock, M.~Kaplinghat, S.~Garrison-Kimmel,
  J.~Onorbe, and L.~A. Moustakas, ``{Cosmological Simulations with
  Self-Interacting Dark Matter I: Constant Density Cores and Substructure},''
  \href{http://dx.doi.org/10.1093/mnras/sts514}{{\em Mon. Not. Roy. Astron.
  Soc.} {\bfseries 430} (2013) 81--104},
\href{http://arxiv.org/abs/1208.3025}{{\ttfamily arXiv:1208.3025
  [astro-ph.CO]}}.
%%CITATION = ARXIV:1208.3025;%%.

\bibitem{Peter:2012jh}
A.~H.~G. Peter, M.~Rocha, J.~S. Bullock, and M.~Kaplinghat, ``{Cosmological
  Simulations with Self-Interacting Dark Matter II: Halo Shapes vs.
  Observations},'' \href{http://dx.doi.org/10.1093/mnras/sts535}{{\em Mon. Not.
  Roy. Astron. Soc.} {\bfseries 430} (2013) 105},
\href{http://arxiv.org/abs/1208.3026}{{\ttfamily arXiv:1208.3026
  [astro-ph.CO]}}.
%%CITATION = ARXIV:1208.3026;%%.

\bibitem{Gavin:1985ph}
S.~Gavin, ``{TRANSPORT COEFFICIENTS IN ULTRARELATIVISTIC HEAVY ION
  COLLISIONS},''
\href{http://dx.doi.org/10.1016/0375-9474(85)90190-3}{{\em Nucl. Phys.}
  {\bfseries A435} (1985) 826--843}.
%%CITATION = NUPHA,A435,826;%%.

\bibitem{Kadam:2015xsa}
G.~P. Kadam and H.~Mishra, ``{Dissipative properties of hot and dense hadronic
  matter in an excluded-volume hadron resonance gas model},''
  \href{http://dx.doi.org/10.1103/PhysRevC.92.035203}{{\em Phys. Rev.}
  {\bfseries C92} no.~3, (2015) 035203},
\href{http://arxiv.org/abs/1506.04613}{{\ttfamily arXiv:1506.04613 [hep-ph]}}.
%%CITATION = ARXIV:1506.04613;%%.

\bibitem{Satapathy:2020sxs}
S.~Satapathy, J.~Dey, P.~Murmu, and S.~Ghosh, ``{On lower bound of relaxation
  time for massless fluid in presence of magnetic field},'' in {\em {64th DAE
  BRNS Symposium on nuclear physics Lucknow, Uttar Pradesh, India, December
  23-27, 2019}}.
\newblock 2020.
\newblock
\href{http://arxiv.org/abs/2001.08609}{{\ttfamily arXiv:2001.08609 [hep-ph]}}.
\newblock
%%CITATION = ARXIV:2001.08609;%%.

\bibitem{Kaplinghat:2015aga}
M.~Kaplinghat, S.~Tulin, and H.-B. Yu, ``{Dark Matter Halos as Particle
  Colliders: Unified Solution to Small-Scale Structure Puzzles from Dwarfs to
  Clusters},'' \href{http://dx.doi.org/10.1103/PhysRevLett.116.041302}{{\em
  Phys. Rev. Lett.} {\bfseries 116} no.~4, (2016) 041302},
\href{http://arxiv.org/abs/1508.03339}{{\ttfamily arXiv:1508.03339
  [astro-ph.CO]}}.
%%CITATION = ARXIV:1508.03339;%%.

\bibitem{Lin:2019uvt}
T.~Lin, ``{Dark matter models and direct detection},''
  \href{http://dx.doi.org/10.22323/1.333.0009}{{\em PoS} {\bfseries 333} (2019)
  009},
\href{http://arxiv.org/abs/1904.07915}{{\ttfamily arXiv:1904.07915 [hep-ph]}}.
%%CITATION = ARXIV:1904.07915;%%.

\bibitem{book:1124099}
C.~C. Chapman~S., Cowling~T.G., {\em The Mathematical Theory of Non-uniform
  Gases: An Account of the Kinetic Theory of Viscosity, Thermal Conduction and
  Diffusion in Gases}.
\newblock Cambridge Mathematical Library. Cambridge University Press, 3ed.~ed.,
  1970.

\bibitem{Ade:2013zuv}
{\bfseries Planck} Collaboration, P.~A.~R. Ade {\em et~al.}, ``{Planck 2013
  results. XVI. Cosmological parameters},''
  \href{http://dx.doi.org/10.1051/0004-6361/201321591}{{\em Astron. Astrophys.}
  {\bfseries 571} (2014) A16},
\href{http://arxiv.org/abs/1303.5076}{{\ttfamily arXiv:1303.5076
  [astro-ph.CO]}}.
%%CITATION = ARXIV:1303.5076;%%.

\bibitem{Farooq:2016zwm}
O.~Farooq, F.~R. Madiyar, S.~Crandall, and B.~Ratra, ``{Hubble Parameter
  Measurement Constraints on the Redshift of the Deceleration–acceleration
  Transition, Dynamical Dark Energy, and Space Curvature},''
  \href{http://dx.doi.org/10.3847/1538-4357/835/1/26}{{\em Astrophys. J.}
  {\bfseries 835} no.~1, (2017) 26},
\href{http://arxiv.org/abs/1607.03537}{{\ttfamily arXiv:1607.03537
  [astro-ph.CO]}}.
%%CITATION = ARXIV:1607.03537;%%.

\bibitem{Amanullah:2010vv}
R.~Amanullah {\em et~al.}, ``{Spectra and Light Curves of Six Type Ia
  Supernovae at 0.511 < z < 1.12 and the Union2 Compilation},''
  \href{http://dx.doi.org/10.1088/0004-637X/716/1/712}{{\em Astrophys. J.}
  {\bfseries 716} (2010) 712--738},
\href{http://arxiv.org/abs/1004.1711}{{\ttfamily arXiv:1004.1711
  [astro-ph.CO]}}.
%%CITATION = ARXIV:1004.1711;%%.

\bibitem{Suzuki:2011hu}
{\bfseries Supernova Cosmology Project} Collaboration, N.~Suzuki {\em et~al.},
  ``{The Hubble Space Telescope Cluster Supernova Survey: V. Improving the Dark
  Energy Constraints Above z>1 and Building an Early-Type-Hosted Supernova
  Sample},'' \href{http://dx.doi.org/10.1088/0004-637X/746/1/85}{{\em
  Astrophys. J.} {\bfseries 746} (2012) 85},
\href{http://arxiv.org/abs/1105.3470}{{\ttfamily arXiv:1105.3470
  [astro-ph.CO]}}.
%%CITATION = ARXIV:1105.3470;%%.

\bibitem{Sahni:2002fz}
V.~Sahni, T.~D. Saini, A.~A. Starobinsky, and U.~Alam, ``{Statefinder: A New
  geometrical diagnostic of dark energy},''
  \href{http://dx.doi.org/10.1134/1.1574831}{{\em JETP Lett.} {\bfseries 77}
  (2003) 201--206}, \href{http://arxiv.org/abs/astro-ph/0201498}{{\ttfamily
  arXiv:astro-ph/0201498 [astro-ph]}}.
[Pisma Zh. Eksp. Teor. Fiz.77,249(2003)].
%%CITATION = ASTRO-PH/0201498;%%.

\end{thebibliography}\endgroup
 \end{document}